\documentclass[superscriptaddress,prd,preprint,nofootinbib]{revtex4}
\usepackage{amsmath,amssymb}
\usepackage{graphicx,epsfig}
\def\simgt{\mathrel{\lower2.5pt\vbox{\lineskip=0pt\baselineskip=0pt
           \hbox{$>$}\hbox{$\sim$}}}}
\def\simlt{\mathrel{\lower2.5pt\vbox{\lineskip=0pt\baselineskip=0pt
           \hbox{$<$}\hbox{$\sim$}}}}

\def\vev#1{\left\langle#1\right\rangle}
\def\beq{\begin{equation}}
\def\eeq{\end{equation}}
\def\beqa{\begin{eqnarray}}
\def\eeqa{\end{eqnarray}}
\begin{document}

\preprint{UCB-PTH-03/34}
\preprint{LBNL-54211}
\preprint{BUHEP-03-22}

\title{CMB Signals of Neutrino Mass Generation}

\author{Z.~Chacko}
\affiliation{Department of Physics, University of California, Berkeley,\\
      and\\
      Theoretical Physics Group, Lawrence Berkeley National Laboratory,\\
      Berkeley, CA 94720, USA}
\author{Lawrence J.~Hall}
\affiliation{Department of Physics, University of California, Berkeley,\\
      and\\
      Theoretical Physics Group, Lawrence Berkeley National Laboratory,\\
      Berkeley, CA 94720, USA}
\author{Takemichi Okui}
\affiliation{Department of Physics, Boston University,\\
     Boston, MA 02215, USA}
\author{Steven J.~Oliver}
\affiliation{Department of Physics, University of California, Berkeley,\\
                          and\\
      Theoretical Physics Group, Lawrence Berkeley National Laboratory,\\
      Berkeley, CA 94720, USA}

\begin{abstract}
We propose signals in the cosmic microwave background to probe the type 
and spectrum of neutrino masses. In theories that have spontaneous breaking 
of approximate lepton flavor symmetries at or below the weak scale, 
light pseudo-Goldstone  bosons recouple to the cosmic neutrinos after 
nucleosynthesis  and affect the acoustic oscillations  of the electron-photon 
fluid during the eV era. Deviations from the Standard Model are predicted for 
both the total energy density in radiation during this epoch, $\Delta N_\nu$, 
and for the multipole of the n'th CMB peak at large n, $\Delta l_n$. The latter 
signal is difficult to reproduce other than by scattering of the known neutrinos, 
and is therefore an ideal test of our class of theories. In many models, the 
large shift $\Delta l_n \approx 8 n_S$ depends on the number of neutrino species 
that scatter via the pseudo-Goldstone  boson interaction. This interaction is 
proportional to the neutrino masses, so that the signal reflects the neutrino 
spectrum. The prediction for $\Delta N_\nu$ is highly model dependent, but can 
be accurately computed within any given model. It is very sensitive to the number 
of pseudo-Goldstone bosons, and therefore to the underlying symmetries of the 
leptons, and is typically in the region of $0.03 < \Delta N_\nu < 1$. This signal 
is significantly larger for Majorana neutrinos than for Dirac neutrinos, and, 
like the scattering signal, varies as the spectrum of neutrinos is changed 
from hierarchical to inverse hierarchical to degenerate.
\end{abstract}

\maketitle
\section{\label{sec:intro} Introduction}

Over the last five years, a variety of experiments, involving
neutrinos from the sun, atmospheric air showers, nuclear reactors and
accelerators, have amassed compelling evidence that neutrinos have
non-zero masses \cite{neutrinomass}. A remarkable feature of this data is
that the two measured leptonic mixing angles are large. This was a
surprise: theories which unify quarks and leptons had led to the
expectation that the mixing amongst leptons, like that between quarks,
would be small. Hence the data has sparked considerable activity
directed toward understanding the origin of these large mixing
angles.  The more fundamental question of why the neutrino masses are
so much smaller than the charged fermion masses has received less
attention. There is a general belief that this problem was solved many
years ago by the seesaw mechanism \cite{seesaw}. Indeed, one sometimes
forgets that the data has not confirmed the seesaw mechanism, and it
is worth stressing that there is no known experiment or observation
which could test this plausible idea. Given our current theoretical
understanding of effective field theories, the seesaw mechanism does
indeed appear to give a very natural explanation for the lightness of
the neutrinos.  But this, like the belief in small mixing angles, is a
theoretical prejudice, and in neutrino physics we have learnt to
expect surprises.

In this paper we pursue an alternative idea: the neutrinos are light
because they are protected from acquiring a mass by a global symmetry
which is not broken until energies at or beneath the weak scale. The
philosophy is precisely opposite to that of the seesaw mechanism ---
the underlying physics is not all at extremely high energies where it is
hard to test, rather some is at very low energies, and we have missed it
because it couples only to neutrinos. Instead of right-handed
neutrinos acquiring very large masses at some high scale of lepton
number breaking, neutrino masses arise from symmetry breaking at much lower
energies. We explore the cosmological consequences of neutrino mass
generation at a phase transition at or below the weak scale. 

While we do not know of
laboratory tests for this idea, signals in the cosmic
microwave background (CMB) could not only answer whether the neutrinos
are Majorana or Dirac, but also distinguish between the 
hierarchical, inverted and degenerate patterns of neutrino 
masses. This signal results because the acoustic oscillations at the
eV era are sensitive to the total energy density in relativistic
particles and to whether these relativistic particles scatter or
free-stream. A measurement of the energy density of this radiation at the few
percent level, and especially its scattering characteristics,
will probe physical processes occurring in the
neutrino fluid at and before the eV era.

\section{Neutrino Mass Generation}
\label{sec:numass}

\subsection{Why are Neutrinos Light?}

The mass scale of all the quarks and charged leptons is set by the
scale $v$ of electroweak symmetry breaking: $\vev{h} = v$, 
where $h$ is the electroweak Higgs field. 
The interaction generating these masses is assumed to
have the form $\overline{\psi}_L \psi_R h$, where $\psi$ can be any of the
quarks or charged leptons. If neutrino masses also originate from such
an interaction, it is hard to understand why neutrinos would be so
much lighter than the charged fermions. The beauty of the seesaw
mechanism is that it explains why, of all the fermions, it is the
neutrinos which are light. The only fermion that does not couple to
the known gauge interactions is the right-handed neutrino, and hence
it is not protected by gauge symmetry from acquiring a large Majorana 
mass, $M_R$. On integrating it out of the theory, the left-handed 
lepton doublet $\ell$ acquires an interaction which is bilinear in $h$
\begin{equation}
\frac{1}{M_R} \ell\ell \; hh,
\label{eq:seesaw1}
\end{equation}
leading to a neutrino mass which is quadratic in $v$, $m_\nu \approx
v^2/M_R$, rather than the linear formula that applies to the charged
fermions: $m \approx v$. Indeed, Eqn. (\ref{eq:seesaw1}) provides a 
more primitive understanding of why the neutrinos are light. As long 
as the low energy effective theory does not contain right-handed neutrinos, 
there are no renormalizable operators that give a neutrino mass --- the first 
neutrino mass operator appears at dimension 5.

We propose instead that the neutrinos are protected from acquiring a mass 
at the weak scale by a global symmetry, as in {\cite{majoron1}}, 
{\cite{majoron2}}.
The operator (\ref{eq:seesaw1}) does not possess such a symmetry, and hence is
not the TeV description of neutrino masses we seek. The operators relevant for
neutrino masses must involve a new scalar field $\phi$ which carries a charge
under the global symmetry. Thus the TeV description of neutrino masses is given
by operators of the form
\begin{equation}
 \ell n \; h  \left({\phi \over M}\right)^N, \;  \; \;  
  \ell\ell \; {hh \over M} \left({\phi \over M}\right)^N,
\label{eq:tevops}
\end{equation}
where $n$ represents the right-handed neutrino, $M$ is a mass scale
larger than $v$, and $N$ is a positive integer. 
The first operator applies if lepton number is
conserved, leading to Dirac neutrino masses, otherwise the second
operator applies and the neutrinos are Majorana. Unlike the case of
the seesaw mechanism, there is no preference for the neutrinos to be
Majorana.  Very stringent bounds would result if the Goldstone
coupled to charged leptons and quarks; but these couplings are 
predicted to be absent because the charged fermion masses are not 
protected by the global symmetry. 

At sufficiently high temperatures in the hot big bang, the $\phi$ and
$n$ particles will be in thermal equilibrium with the particles of the
standard model. However, if $\phi$ and $n$ only interact with standard
model particles via (\ref{eq:tevops}), then they will drop out of
thermal equilibrium as the universe cools so that there is an era of
two separate sectors. During this era we assume that sufficient
entropy is created in the standard model sector, from phase transitions
or from heavy particle annihilations, so that the temperature rises
significantly above that of the $(\phi,n)$ sector. Hence big bang
nucleosynthesis is essentially unchanged by the extra sector. 

Below we describe a minimal model for the flavor
symmetry breaking sector. However, the details of any particular model
are not as important as the model independent mechanism. Once symmetry
breaking occurs in the $\phi$ sector, a set of Goldstone bosons, $G$,
are produced. The CMB signals result from the interactions of $G$
with neutrinos at very low energies, and will be discussed in section III. 

\subsection{A Minimal Model: $U(1)_L$}

We choose the global symmetry to be lepton number, $U(1)_L$, defined
with charge +1 on all neutrino fields ({\it i.e.} on both $\nu_i$ and $n_i$).
Just below the scale of electroweak symmetry breaking the neutrino mass 
generation sector is described by
\begin{equation}
\mathcal{L}_\nu 
 = \sqrt{2} g_i 
     \left( \nu_i n_i \phi, \; \frac{1}{ 2} \nu_i \nu_i \phi \right) 
      + \mbox{h.c.}
   - (- \mu^2 \phi^\dagger \phi + \frac{\lambda}{2} (\phi^\dagger \phi)^2)
\label{eq:minmod}
\end{equation}
together with kinetic energy terms for $\nu_i, \phi$ (and for $n_i$ if the
neutrino is Dirac). For simplicity we have taken the neutrino interaction
to be linear in $\phi$ by requiring
the complex scalar $\phi$ to have lepton number $-2$.
The index $i$ runs over the three generations of
neutrinos, and we have rotated the neutrino fields to a mass
eigenstate basis. Small values for the dimensionless couplings $g_i$ 
are perfectly natural, reflecting the hierarchy between $v$ and $M$.

We have taken the sign of the scalar mass term
negative to ensure that $U(1)_L$ is
spontaneously broken, $\vev{\phi} = f/\sqrt{2} = \mu/ \sqrt{\lambda}$.
This minimal model does not address the origin of the neutrino mass
ratios, which follow from $g_{1,2}/g_3$. We assume the largest
coupling, $g_3$, does not involve any small dimensionless parameter,
so that $g_3$ is $v/M$  for the Dirac case or $(v/M)^2$ for the
Majorana case, giving a neutrino mass
$m_{\nu_3}$ of $(v/M)f$ or $(v/M)^2 f$ respectively.
The mass scale $M$ is then
$(f/m_{\nu_3})v$ or $\sqrt{f/m_{\nu_3}} \; v$, respectively. This should be
compared with the seesaw result: $M_R \approx (v/m_{\nu_3})v \approx
10^{12}v$. For $f \ll v$, the scale of the underlying 
physics is reduced: $M \ll M_R$. At what scale, $f$, should the global symmetry
be broken? If we take $f$
all the way down to $m_{\nu_3}$, then $M \approx v$ so that the physics
generating the non-renormalizable operators (\ref{eq:tevops}) becomes
accessible to high energy colliders. However, in this case the
dimensionless coupling $g_3$ is of order unity, so that in the early universe
both $\phi$ and $n$ become part of the thermal bath
during the MeV era, conflicting with big bang nucleosynthesis \cite{BBN}. 
If $f$ were larger than the electroweak scale, then $g_3 \simlt 10^{-12}$,
which is too small to generate the CMB signals we have in mind. 
Hence we study an intermediate situation where 
\begin{equation}
m_\nu \ll f \simlt v.
\label{eq:f}
\end{equation}
We will construct theories in which such low symmetry breaking scales arise
naturally in section \ref{sec:susy}. Breaking lepton number below the weak
scale also avoids the potential danger that the baryon asymmetry will be
erased {\cite{Olive}}.

In the spontaneously broken phase, $\phi = (f + H+iG)/ \sqrt{2}$, 
where $G$ is a physical Goldstone boson, and $H$ a
Higgs boson. The coupling of the neutrino to $G$ and $H$ is given by
\begin{equation}
\mathcal{L}_\nu = g_i \left( \nu_i n_i,  \;
                      \frac{1}{ 2} \nu_i \nu_i \right) (H+iG).
\label{eq:GHint}
\end{equation}

The analysis of this paper is almost entirely based on these
couplings, and the symmetry breaking sector that leads to them is of
lesser importance. The masses of $G$ and $H$ play a very important
role. If the self interactions of $\phi$ are of order unity, then one
expects $m_H \approx f$. In this case it is $G$ which is of interest
to us. We assume that $G$ is actually a pseudo-Goldstone boson, 
with a non-zero mass $m_G$, as studied in {\cite{massivemajoron}}. If this
mass arises from a dimension 5 interaction suppressed by the Planck
scale, $M_P$, then we expect $m_G^2 \approx f^3/M_P$. We will not
limit ourselves to this case, but take $m_G \ll f$ to be a free
parameter.  If $\phi$ is weakly coupled so that $m_H \ll f$, then the
Higgs can also play an important role in generating CMB
signals. However, Higgs particles lighter than $f$ require further
small parameters.  In this paper we focus on signals from
pseudo-Goldstone bosons (PGBs). They are naturally light, and they
interact with neutrinos only via the couplings $g_i$.  Furthermore,
they generically have interactions among themselves with strength
proportional to explicit symmetry breaking that gives PGBs mass.  In
more general models of spontaneously broken lepton flavor symmetry
there are many PGBs, $G_A$.

\subsection{Origins of CMB Signals}

The interactions of PGBs with neutrinos can alter the energy density
in neutrinos and cause neutrinos to scatter
rather than free stream during the eV era. Both effects leave
characteristic signals in the CMB.  The PGBs, $G_A$, interact only via
the small dimensionless couplings $g_i$ to neutrinos, implying that in
the early universe the rate for neutrinos to scatter into $G_A$ has a
recoupling form; that is, the rate increases relative to the expansion
rate as the temperature, $T$, drops.  At some recoupling temperature
some subset of the $n_i,G_A$ sector recouples to the left-handed
neutrinos, $\nu_i$, so that this subset gets reheated. Different subsets
may get reheated at various recoupling temperatures. However, while
this reheating creates entropy, it does not change the total radiation
energy density, so recoupling itself does not lead to a change in the
radiation energy density at the eV era. 

As the $\nu,n,G$ fluid cools, the temperature will drop beneath the
mass of one of the PGBs, $G_H$. Since $G_H$ is in thermal equilibrium
with $\nu,n$ and $G_A$, its number density becomes exponentially
reduced via decays or annihilations and the neutrino fluid is
adiabatically heated, which {\em does} lead to a change in the
radiation energy density.  The size of this signal depends on how many
right-handed neutrinos and scalars are recoupled, which depends on
whether the neutrinos are Dirac or Majorana and on the strength of the
couplings $g_i$ that each neutrino has with $G_A$.  Provided $G_H$ has
recoupled and then disappeared before the eV era, the cosmic microwave
background will have acoustic oscillations which reflect a radiation
energy density that differs from standard cosmology and depends on the
type and spectrum of the neutrinos.

If a light PGB recouples to neutrinos before the eV era, but has a mass
less than 1 eV, it may prevent one or more of the neutrino species
from free-streaming. In the standard model, neutrino free streaming
shifts the multipoles of the nth CMB peak by a large amount, $\Delta l
\approx -23$, at large $l$, so the absence of free-streaming will produce a 
large signal in future CMB datasets.

\section{The CMB Signals.}
\label{sec:cmbsignals}

In the previous section we have discussed an alternative origin for
light neutrino masses, involving extra states at low energy, including
Goldstone bosons, Higgs and possibly right-handed neutrino states. We
have outlined how the interaction of these extra states could lead to
signals in the CMB. In this section we study each of the CMB signals
in some depth. We give general formulas for both signals in a model
independent way, and discuss the range of new physics which could lead
to each signal.

\subsection{\label{subsec:rhorelsig} The Relativistic Energy Density Signal}

Measurements of the precise form of the CMB acoustic oscillations
provide a powerful constraint on the relativistic energy density of
the universe during the eV era, $\rho_{rel}$. In the standard model,
$\rho_{rel}$ is precisely predicted during this era, so these
measurements are powerful probes of non-standard physics.  Increasing
$\rho_{rel}$ has several physical effects. For example,  last scatter
occurs at a fixed temperature, and hence as  $\rho_{rel}$ increases so
the time at last scatter decreases. This decreases the horizon at last
scatter and hence shifts the acoustic peaks to higher multipole
$l$. An increase in  $\rho_{rel}$ leads to a  lowering of the redshift
of matter radiation equality. This leads to larger amplitude acoustic peaks
at low $l$ \cite{Hu:1995fq}, and a marked increase in the
damping of the peaks at large $l$ \cite{BS}.

What mechanism would allow the total radiation energy density during
the eV era to differ from that predicted by standard cosmology? 
We begin by discussing three important types of process: fragmentation, 
recoupling and disappearance by decay or annihilation.

In cosmology it is well known that particles which interact with
each other at very high temperature may no longer have thermal
communication at lower temperature. In general one should expect that
as the universe cools the fluid fragments into multiple components or
sectors. This fragmentation occurs whenever there are no large
renormalizable interactions between particles. Neutrinos provide the 
most familiar example: below the weak scale they only interact via the
non-renormalizable Fermi interaction and they fragment away from the
electron/photon fluid at the MeV era. It appears quite likely that
dark matter and dark energy are sectors that fragmented from the
visible sector at some stage of cosmological evolution. We therefore
write the relativistic energy density after fragmentation of the known 
neutrinos as
\begin{equation}
\rho_{rel} = \frac{\pi^2}{ 30} \left( 2 T^4 + g_\nu T_\nu^4 
              + \sum_a g_a T_a^4 \right)
\label{eq:rhorad}
\end{equation}
where $T,T_\nu,T_a$ are the temperatures of the photons, neutrinos and
other sectors and the spin degeneracies $g_\nu, g_a$ for neutrinos and
other radiation sectors include a factor of 7/8 for fermions. It is
not always necessary that each sector actually be in thermal
equilibrium. For example, during this era in the standard cosmology
the neutrinos are free-streaming.  In the standard cosmology, $g_a =
0$, $g_\nu=21/4$ and $T_\nu = (4/11)^{1/3} T$, so that
\begin{equation}
\rho_{rel} = \frac{\pi^2}{ 30} (2 + 0.45 N_\nu) T^4
\label{eq:rhorad2}
\end{equation}
where the number of neutrinos, $N_\nu$, is 3. In non-standard
cosmologies we use (\ref{eq:rhorad2}) to define $N_\nu$, so that
$N_\nu$ may differ from 3 even if there are three species of neutrinos.

It is important to distinguish two very different ways in which CMB
experiments could measure $N_\nu \neq 3$.  If $N_\nu$ does not change
between nucleosynthesis (BBN) and the eV era (CMB), then CMB could
then discover $\delta N_\nu = \pm (0.5 - 1.0)$, depending on the
uncertainties from BBN. Such a signal could be very significant given
the small uncertainties possible in future CMB measurements. Since
$N_\nu = (\rho_\nu + \sum_a \rho_a)/ 0.23\rho_\gamma$, this could
probe the radiation density $\rho_a$ in some sector that fragmented
from the standard model sector before, perhaps much before, BBN.

Perhaps less well known than the phenomenon of fragmentation is that of
recoupling. If the renormalizable couplings between particles in 
different fragmented sectors are small rather than vanishing,
then eventually, as the universe cools, the
sectors will recouple back into a single thermal component. The
smaller the renormalizable coupling the lower the recoupling
temperature.

In this paper we will be concerned with recoupling contributing to a signal
arising from a change in the ratio $(\rho_\nu + \sum_a \rho_a)/ \rho_\gamma$
between BBN and CMB \cite{BBNCMB}. We find such a possibility particularly
exciting because it indicates that new physics is affecting cosmological
evolution after BBN. It could arise from a non-standard evolution of
$\rho_\gamma$ or of ($\rho_\nu + \sum_a \rho_a$). Particularly large effects
result if the electron/photon fluid recouples to some previously fragmented
sector of spin degeneracy $g$. If the temperature of the fragmented sector
is small prior to recoupling, the photon cooling results in a large relative
increase in the importance of the neutrinos, giving $\delta N_\nu = 3.7 g$.
This has already been excluded by the first year of the WMAP data
{\cite{WMAP}}, except for the case of $g=1$
\cite{BBN}\cite{Hannestad:2003xv}. Photon heating is also possible, for
example from the out-of-equilibrium decay of a non-relativistic species.

A particularly important way of changing the radiation density of any
sector is if some particle species in that sector becomes
non-relativistic. If the number density of the heavy particle
maintains an exponentially suppressed equilibrium form, then the decay
or annihilation of the particle occurs at constant entropy.  This results
in an increase in the temperature of the remaining radiation of that
sector by a factor $(g_i/g_f)^{1/3}$, where $g_i$ and $g_f$ are spin
degeneracies of the radiation of that sector before and after the
disappearance of the heavy species. This mechanism is familiar from
the annihilation of electron/positron pairs which heat the photons
giving $T_\nu = (4/11)^{1/3} T_\gamma$.  As an example of a
non-standard evolution of $\rho_\gamma$, suppose that the photon
first recouples to a sector of spin degeneracy $g$, and then all the
species of that sector become non-relativistic before the eV era, the
CMB signal is $\delta N_\nu = 3(2/(2+g))^{1/3} -3$. A wide range of
$g$ is allowed by the WMAP data and could be observed by future CMB
experiments.

In this paper, we will be interested in the case that a CMB signal
arises because of a non-standard evolution of $\rho_\nu + \sum_a
\rho_a$ after BBN, even though the signals tend to be smaller than
those which arise from a non-standard evolution of $\rho_\gamma$.
Such a signal could arise whenever some particle of a fragmented
sector becomes non-relativistic and disappears. While the temperature
of this sector would rise by a factor of $(g_i/g_f)^{1/3}$, this
typically does not give an observable signal since the energy density
in the sector is highly sub-dominant. We will explore theories of
neutrino mass generation where the sector that generates the neutrino
masses, including the right-handed neutrinos if they are Dirac,
recouples to the sector of the left-handed neutrinos. This recoupling
ensures that the physics of the new sector is not sub-dominant. While
the recoupling of two such relativistic sectors does not by itself lead to a
change in the total energy density\footnote{When heat is exchanged
between two relativistic sectors, the form of the energy momentum
tensor is not changed, hence the expansion rate of the universe is
also unchanged. Since there is no change in the gravitational energy,
the total fluid energy is unchanged.}, the effects of particle decay
or annihilation in the new sector does give signals which depend on the
type and spectrum of the neutrinos, and on the mass generation
mechanism, and which can be computed from equilibrium thermodynamics.

Suppose that after BBN $n_R$ of the left-handed neutrino species
recouple to some sector of $G,n$ particles with spin degeneracy
$g$. This recoupling may occur at a variety of temperatures, but we
assume that at some stage the resulting $\nu,G,n$ fluid thermalizes at
a single temperature. Subsequently we assume that some subset of the
states of this recoupled sector, with spin degeneracy $g_H$, become
heavy. As the universe expands further, but well before the eV era,
the number density of the heavy states becomes exponentially
suppressed so that their entropy is transferred to the lighter
states. This may occur at several stages with different heavy species
having different masses. If the whole process occurs without chemical
potentials, then the prediction for the relativistic energy
density during the eV era is then given by
\begin{equation}
N_\nu =  3 - n_R + n_R \left( 
\frac{n_R +\frac{4}{ 7} g }{ n_R + \frac{4}{ 7} (g - g_H)} \right)^{1/3}.
\label{eq:Nnuenergy}
\end{equation}

In the case where the recoupling reactions lead to chemical potentials
this prediction is modified.  It is no longer sufficient to
calculate the neutrino temperature after the pseudo-Goldstone bosons
disappear.  Instead, both the temperature and all chemical potentials
must be determined at each step.  Typically all chemical potentials
are related such that there exists only one additional degree of
freedom.

At the first step, when the pseudo-Goldstone bosons equilibrate, the
total energy density is conserved.  In addition, the presence of a
non-zero chemical potential implies that there exists an additional
conserved charge.  These two conservation laws provide us with two
equations to solve for the two unknowns: temperature and chemical
potential.  When the Goldstone bosons disappear energy is not
conserved.  Instead comoving entropy density, together with the
additional quantum number, are conserved.  This again allows us to
calculate the final temperature and chemical potential.  From these
two parameters we are able to calculate the final energy density in
neutrinos and finally the effective number of neutrinos at
matter-radiation equality
 \beq 
   N_\nu = (3-n_R)+n_R \frac{\rho_{\nu_R}(T,\mu)}{\rho_{\nu_R}(T_{SM},0)} 
\label{eq:Nnumu} 
\eeq 
where $T_{SM}= (4/11)^{1/3}T_{\gamma}$ is the temperature the neutrinos
would have had in the standard model and $\rho_{\nu_R}$ is the energy density
in the $n_R$ neutrino species that recoupled.

\subsection{The Neutrino Scattering Signal}

Recently Bashinsky and Seljak have given an analytic understanding of
the effects of neutrino free-streaming on the position and amplitudes
of the CMB acoustic peaks in the standard model \cite{BS}. Here we
briefly summarize some of their results, using their notation.

In the conformal Newtonian gauge, the Robertson-Walker metric with scalar 
metric perturbations takes the form
\begin{equation}
ds^2 = a^2(\tau) \left( -(1+2 \Phi) d \tau^2 + (1-2 \Psi) d {\bf r}^2
\right),
\label{eq:metric}
\end{equation}
where $\tau$ is conformal time and $a$ is the cosmological scale
factor. In this gauge, the density perturbation in the relative photon
number density $d_\gamma = \delta n_\gamma({\bf r}) /
n_\gamma({\bf r})$ satisfies the equation
\begin{equation}
\frac{d^2 d_\gamma}{ d \tau^2} - \frac{1}{ 3} \nabla^2 d_\gamma =
\nabla^2 (\Psi + \Phi).
\label{eq:gammapert}
\end{equation}
In the absence of any particle species which free-streams, the 
energy-momentum tensor takes the form for a locally isotropic fluid,
resulting in the equality of the scalar metric perturbations: $\Psi =
\Phi$. However, neutrino free-streaming introduces a direction at each
locality, so that the energy-momentum tensor becomes anisotropic, with
off-diagonal entries in the spatial subspace proportional to a
free-streaming potential $\pi_\nu$. This anisotropy induces a
difference between the scalar metric perturbations:
\begin{equation}
\Psi - \Phi = 6 \frac {R_\nu \pi_\nu}{ \tau^2},
\label{eq:perturbdiff}
\end{equation}
where $R_\nu$ is the fraction of radiation in neutrinos: 
\begin{equation}
R_\nu =  \frac{\rho_\nu}{ \rho_\nu + \rho_\gamma} \simeq
 \frac {0.23 N_\nu}{1 +  0.23 N_\nu}.
\label{eq:Rnu}
\end{equation}

If the initial density perturbations are adiabatic, then the
perturbation in the relative number density of all species, before
they enter the horizon, are given by $d({\bf r}) = -3 \xi({\bf r})$,
where $ \xi(\bf{r})$ describes the primordial perturbation. Solving
first for $\pi_\nu$, next for the metric perturbations, and finally
for the photon perturbation at comoving wavenumber $k$, Bashinsky and
Seljak find acoustic oscillations in the radiation dominated era for
high $k$ of the form
\begin{equation}
d_\gamma(\tau,k) = 3  \xi(k) (1 + \Delta_\gamma) \cos (k \tau /
\sqrt{3} + \delta \varphi)
\label{eq:pertsol}
\end{equation}
where, to leading order in $R_\nu$, $\Delta_\gamma = - 0.27 R_\nu$ and
\begin{equation}
\delta \varphi =  0.19 \pi R_\nu.
\label{eq:deltaphi}
\end{equation}
The amplitude shift of the primordial spectrum, $\Delta_\gamma$, can
only be probed by observations which compare the photon and cold dark
matter perturbations, and we do not consider them further.
The $n$th peak of the CMB acoustic oscillations occurs at a wavenumber
$k_n$ such that the mode has $n$ half cycles of oscillation between
horizon crossing and last scatter (LS):
\begin{equation}
 \frac{k_n \tau_{LS}}{ \sqrt{3}} = n \pi - \delta \varphi.
\label{eq:k_n}
\end{equation}
Since the multipole of the peak $l_n \propto k_n$, $\delta \varphi$
causes a shift in the position of the $n$th peak. This analytic
solution accurately reproduces the numerical results obtained by the
CMBFAST code, and it is apparent that this shift of the position of the
peaks is purely due to the free-streaming of the neutrinos, since, in the
absence of free-streaming, both $\pi_\nu$ and $\delta \varphi$
vanish. Thus, the free-streaming of $N_\nu$ species of neutrinos shifts
the positions of the peaks, at large $n$, by
\begin{equation}
\Delta l_n \simeq - 57 \left( \frac{0.23 N_\nu}{ 1+ 0.23 N_\nu} \right)
\left(  \frac{\Delta l_{peak}}{ 300} \right),
\label{eq:lshift}
\end{equation}
giving $\Delta l_n \simeq -23.3$ for $N_\nu = 3$.
Here $\Delta l_{peak}$ is the difference in multipole between successive 
peaks at large $n$.
As the number of free-streaming neutrinos is increased above the
standard model value of 3,  so the peaks shift to lower $l$. The
beauty of this signal is that, for adiabatic perturbations, it is not
mimicked by changing other parameters of the theory. While other
parameters do cause a shift in the positions of the peaks, only by
changing the free-streaming behavior can one obtain a non-isotropic
energy-momentum tensor which leads to a shift in the $n$th peak which
is {\em independent} of $n$.

With these results in hand, we immediately see that there is an
important signal in the position of the CMB peaks for theories
where one or more neutrinos are scattering during the eV
era rather than free-streaming. In general one must replace
(\ref{eq:Rnu}) by
\begin{equation}
R_\nu = \frac{\rho_{FS} }{ \rho_{rel}} = \frac{0.23 N_{\nu FS}}{ 1+ 0.23 N_\nu}
\label{eq:Rnugeneral}
\end{equation}
where $\rho_{FS}$ is the energy density of the relativistic components
that free-stream, while $\rho_{rel}$ is the total relativistic energy
density, including both free-streaming and scattering components.
$N_{\nu FS}$ is the energy density of the relativistic free-streaming
particles, expressed as an equivalent number of neutrino species.

Consider a simple limit where the energy density of the free-streaming
neutrinos and the total radiation is standard.  If only $n_{FS}$ of
the three neutrino species free-stream, then
\begin{equation}
\Delta l_n \simeq - 7.8 \; n_{FS} \left(  \frac{\Delta l_{peak}}{ 300} \right).
\label{eq:lshiftscatt}
\end{equation}
Relative to the position of the peaks expected for the standard model
with three free-streaming neutrinos, there is a uniform shift in the
position of the peaks to {\em larger} $l$. This  can only result if some of
the known neutrinos are not free-streaming, and in this simple example
the shift in $l$ is 7.8 for each scattering neutrino species.

The result (\ref{eq:lshiftscatt}) applies if the energy densities are
standard. However, since non-standard physics is
required to prevent free-streaming, it could well be that the neutrino
energy densities are also non-standard, or there could be energy
density from light PGBs. How does this effect the
change in the position of the peaks induced by the  phase shift
$\delta \phi$? From (\ref{eq:Rnugeneral}) we need expressions for
$N_{\nu FS}$ and $N_\nu$. The value of $N_\nu$ is predicted in our
theories by (\ref{eq:Nnuenergy}), or (\ref{eq:Nnumu}) for a non-zero 
chemical potential, and a similar result can be derived for $N_{\nu FS}$.
In terms of these quantities
\begin{equation}
\Delta l_n \simeq - 57 \left( \frac{0.23 N_{\nu FS}}{ 1+ 0.23 N_\nu} \right)
\left(  \frac{\Delta l_{peak}}{ 300} \right).
\label{eq:lshiftdeltaE}
\end{equation}
%

Clearly, in general the position of
the peaks could be determined by a combination of extra neutrinos
(\ref{eq:lshift}), scattering of the known neutrinos
(\ref{eq:lshiftscatt}), and non-standard energy densities of the
known neutrinos (\ref{eq:lshiftdeltaE}).  In this paper we limit our
consideration to the case of theories with three neutrinos. We will
find that the mass generation mechanism gives large regions 
where the prediction (\ref{eq:lshiftdeltaE}) differs significantly
from the shift of $-23.3$ expected in the standard model. The largest
effect comes from $n_{FS} \neq 3$, but a significant deviation from
(\ref{eq:lshiftscatt}) may result because the energy density in each 
free-streaming neutrino is non-standard.

In this section we have discussed two different CMB signals  that
occur in theories with three neutrino species if their interactions are
non-standard. The first probes non-standard neutrino energy densities
which lead to effects such as delayed matter radiation equality and
enhanced damping of acoustic oscillations at higher $l$. However, we have 
shown that such signals can result from a
variety of underlying physics origins which lead to a change of
$N_\nu$. In contrast, the second signal --- a uniform shift of the CMB
peaks to larger $l$, (\ref{eq:lshiftscatt}) --- is a unique signal for
the absence of free-streaming of one or more of the known neutrino
species. 
If the physics leading to neutrino scattering also substantially affects
the neutrino energy densities, then a measurement of $\Delta l_n$ would
not only reveal neutrino scattering, but would also confirm non-standard 
energy densities.

\section{\label{sec:1nu} Signal Regions For One Neutrino}


A complete analysis of the CMB energy density and scattering signals
from the interactions of PGBs with neutrinos is complicated. Several
stages of spontaneous breaking of lepton flavor symmetries can each lead 
to several PGBs, and the combination of neutrino and PGB mass matrices
leads to many parameters. In this section we study the simplest
situation that leads to our CMB signals: a single PGB coupled to a
single neutrino. There are three independent parameters, the PGB and
neutrino masses, $m_G$ and $m_\nu$, and the coupling strength $g$ of
the interaction between them. This coupling parameter $g$ can be traded 
for the symmetry breaking scale at which the PGB is produced $f= m_\nu/g$. 
We aim to discover whether, for typical values of the neutrino mass 
suggested by data, there are large regions in the $(f, m_G)$ parameter
space that give observable CMB signals.
We will first consider the case that
the neutrinos are Majorana, and later discuss the minor changes that 
must be made in the case of Dirac neutrinos.

\subsection{\label{subsec:majorana} Majorana Neutrinos}

Since $G$ is a pseudo-Goldstone boson, at low energies it does not
possess significant self-interactions, so that the only interaction of
interest is given by the Lagrangian term 
 \beq 
   \mathcal{L} \supset\frac{g}{2} \nu \nu (H+iG).  
 \eeq 
This interaction results in three possibly interesting processes: $\nu
\nu \leftrightarrow G$, $\nu \bar{\nu} \leftrightarrow G G$ and $\nu
\nu \leftrightarrow \nu \nu$. We will show that the first process has
a rate that increases relative to the Hubble expansion rate as the
temperature of the universe decreases.  Therefore, this rate may lead
to the recoupling required to produce the signals we seek.  The second
process may also couple neutrinos to the pseudo-Goldstone boson.
However, for $T<\mathrm{MeV}<m_{H}$ the rate has a decoupling form and
is therefore unable to produce either signal.  Instead, demanding
that neutrinos and Goldstone bosons be decoupled prior to big bang
nucleosynthesis will lead to an upper bound on the coupling constant
$g$, and therefore a lower bound on $f$.  
Finally, the rate for $\nu\nu \leftrightarrow \nu\nu$ will be
shown to be too slow to produce a signal.  

The first process, $\nu \nu \leftrightarrow G$, is able to generate
either a $\Delta N_\nu$ and $\Delta l$ signal.  The regions of parameter
space in which each signal is possible are clearly separated by simple
kinematics.  For a change in the effective number of neutrinos $N_\nu$
to be possible, $G$ must go out of the bath prior to $T_{eq} \sim 1$
eV.  Therefore, the scalar mass must satisfy $m_G > 1$ eV.  On the
other hand, for the neutrinos to not free-stream during the era probed
by the acoustic oscillations they must be scattering during the eV
era.  For this to be possible the scalar mass must satisfy $m_G < 1$
eV. 

To see more precisely in what regions of parameter space a signal
results, we will consider each process separately, beginning with the
two-to-one process $\nu \nu \leftrightarrow G$.  This process occurs
at a rapid rate in the presence of a thermal bath of particles despite
the severe kinematic restrictions. For $T \gg m_G>2m_\nu$ the rate is
approximately given by\footnote{The approximations we made here
include: (a) neglecting the difference between $T^n$ and the average
of $E^n$, (b) using simply $T^2/M_{pl}$ and $T^3$ for the expansion
rate $H$ and the number densities, respectively, neglecting prefactors
such as $g\pi^2/30$ and subdominant corrections due to possible
non-zero chemical potentials.}
 \beq 
   \Gamma (\nu \nu \leftrightarrow G) 
      \approx \frac {m_\nu^2 m_G^2}{16\pi f^2 T}.  
 \eeq 
Defining the recoupling temperature as the temperature at which
$\Gamma(T_{rec})=H(T_{rec})$, we find that
 \beq 
   T_{rec}(\nu \nu \leftrightarrow G) \approx 
 \left( \frac{m_\nu^2 m_G^2 M_{pl}}{16\pi f^2} \right)^{1/3}.  
 \eeq 
For temperatures below $T_{rec}$, $G$ will be in
equilibrium with the neutrinos.  If this is the only process that brings
the $G$ into thermal contact with the neutrinos,
then there is a conserved quantity, $n_\nu + 2n_G$, that is left
unchanged as the Goldstone boson comes into equilibrium.  This
conservation law implies the presence of a chemical potential
satisfying $2\mu_\nu = 2\mu_{\bar{\nu}} = \mu_G$.

\begin{figure}
\begin{center}
  \includegraphics[width=16cm]{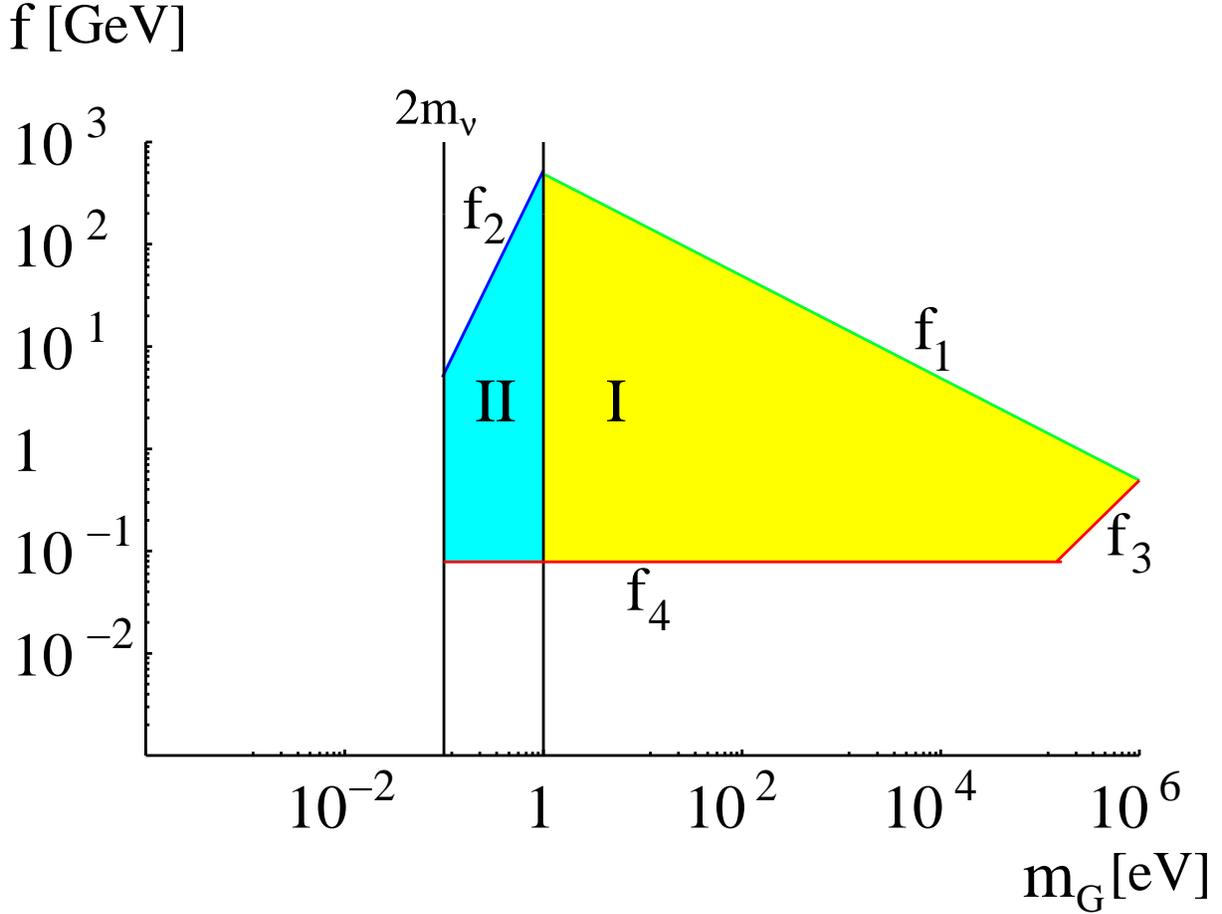}
\caption{Signal regions and cosmological bounds for a single Majorana
  neutrino coupled to a single pseudo-Goldstone boson. The lines and
  regions are labeled as in the text. CMB signals occur throughout 
  the two shaded regions. The area below $f_3$ and $f_4$ is
  excluded by BBN, and in the region above $f_1$ and $f_2$ the
  PGB is too weakly coupled to give any signal. There is an energy
  density signal in region I and a scattering signal in
  region II. We have assumed $m_\nu = 0.05$ eV $\lambda = 1$.}
 \label{fig:majorana}
\end{center}
\end{figure}

If the recoupling temperature is below $m_G$, then the number of
Goldstone bosons produced will be exponentially suppressed and they
will not be able to generate a signal.  Thus we must demand that
$T_{rec} > m_G$.  This leads to a bound on the symmetry breaking scale $f$
 \beq 
  f < f_1= m_\nu \left( \frac{M_{pl}}{16 \pi m_G} \right)^{1/2}.
 \label{eqn:2to1recouplimit}
 \eeq If this limit is satisfied, and $m_G > 1$ eV, then the Goldstone
boson will decay prior to $T_{eq}$.  This will alter the energy
density in neutrinos as discussed in Sec. \ref{subsec:rhorelsig},
changing the effective number of neutrinos, $N_\nu$, that will be
measured by a CMB experiment.  The presence of a non-zero chemical
potential does not alter this key result but will affect the magnitude
of this shift.  The line $f=f_1$ is displayed in figure
\ref{fig:majorana}.

On the other hand, if $m_G < 1$ eV, then $\nu \nu \leftrightarrow G$
will be kinematically allowed during acoustic oscillation and may
therefore prevent the neutrinos from free streaming during this
period.  However, the scattering angle is kinematically
restricted to be quite small, $\theta \sim m_G/T$.  For this process to
isotropize the neutrino momentum they must participate in $N$
scatterings such that $\theta_{tot}^2 \approx (m_G/T)^2 N \approx 1$.
In this case, we must demand that $\Gamma > H N$ resulting in the
limit
 \beq
   f < f_2= m_\nu \left( \frac{M_{pl} m_G^4}{16 \pi T_*^5} \right)^{1/2},
 \label{eqn:2to1scatterlimit}
 \eeq
where $T_* \sim 1$ eV is the temperature at which the perturbation
enters the horizon.

We must also require that $G$ does not come
into equilibrium with the neutrinos prior to the decoupling of the
weak interactions.  If this were to occur they would increase the
energy density in radiation during big bang nucleosynthesis
conflicting with observations of primordial elemental abundances.
Therefore we must demand that $T_{rec} < T_W \sim 1$ MeV.  This places
a bound on $f$ of
 \beq 
   f > f_3 = m_\nu \left( \frac{m_G^2 M_{pl}}{16 \pi T_W^3} \right)^{1/2}.
 \label{eqn:2to1BBNbound}
 \eeq

We now turn to the two-to-two process $\nu \bar{\nu}
\leftrightarrow GG$. In the non-derivatively coupled basis in which we
work, this process is dominated by the exchange of a virtual Higgs
boson for $T>m_\nu$.  The rate is given approximately by
 \beq
   \Gamma(\nu \bar{\nu} \leftrightarrow GG) \approx 
     \frac{m_\nu^2 T^3 m_H^4}{32\pi f^4 (T^2 - m_H^2)^2}.
 \label{eq:2to2rate}
 \eeq

For temperatures below the Higgs mass, $m_H$, this rate goes like $T^3$
and has a decoupling form. Therefore, this process cannot be used to
recouple the Goldstone boson to the neutrinos.  Instead, we must
demand that this process not keep neutrinos and Goldstone bosons in
equilibrium prior to BBN. To insure that the neutrino and Goldstone
boson sectors are decoupled it suffices to demand that they not be in
thermal contact at $T \sim m_H$ when the rate relative to the Hubble
expansion rate is maximal.  The requirement that
$\Gamma(T=m_H) < H(T=m_H)$ leads to a lower bound on $f$ given by
 \beq
   f > \left( \frac{m_H m_\nu^2 M_{pl}}{32 \pi } \right)^{1/4}.
 \eeq
Setting $m_H = \sqrt{\lambda} f$ gives
 \beq
   f > f_4 = \left(  \frac{\sqrt{\lambda} m_\nu ^2 M_{pl}}{32 \pi } \right)^{1/3},
 \eeq
where $\lambda$ is a dimensionless parameter describing the Higgs self coupling. 

%

Finally, we consider $\nu \nu \leftrightarrow \nu \nu$ which is mediated
by the exchange of a virtual Goldstone boson. 
The rate for this process is given approximately by
 \beq
   \Gamma(\nu \nu \leftrightarrow \nu \nu) \approx \frac{m_\nu^4 T^5}{16\pi f^4
 (T^2-m_G^2)^2}.
 \eeq
For our purposes, this process is only interesting if it is able to
prevent the neutrino from free-streaming during the eV era.  This will
be the case if $\Gamma(T\sim 1~\mathrm{eV}) > H(T\sim 1 ~\mathrm{eV})$.
For $m_G < 1$ eV this implies
 \beq
   f < m_\nu  \left( \frac{M_{pl}}{ 16 \pi \mathrm{eV}} \right)^{1/4}.
 \eeq
For $m_\nu < 1$ eV, as implied by the recent WMAP data combined with
the measurements of large scale structure, this bound is larger than
the upper bound coming from $\nu\nu\leftrightarrow GG$.  If $m_G > 1$
eV then the rate will be further suppressed by $\mathrm{eV}^4/m_G^4$.
Therefore, as stated at the beginning of this section, the
$\nu\nu\leftrightarrow \nu\nu$ process is unable to produce an
interesting signal.



All of the above signal regions, together with the cosmological bounds
arising from big bang nucleosynthesis, are displayed in Figure
\ref{fig:majorana}. 
Note that the line for $f_4$ has been drawn for a self coupling 
parameter $\lambda = 1$. For smaller values of $\lambda$ the allowed 
region grows to include lower values of $f$. 
There we see that there are two distinct signal
regions.  In region $I$, $\Delta N_\nu$, as measured from the
relativistic energy density, is non-zero.  Further, the neutrinos (and
Goldstone bosons prior to decay) have a non-zero chemical potential.
In region $II$, $\rho_{rel}$ is unchanged, but
the neutrino is no longer able to free-stream at $T \sim 1$ eV.  As a
result, there will be an overall phase shift in the angular power
spectrum relative to the standard model prediction.

It is important to also understand if there are any bounds on our
scenario from astrophysical processes. Because the pseudo-Goldstones
couple to the quarks and leptons that make up astrophysical objects
only through neutrinos, their effects only show up in highly dense
regions like the cores of supernovae {\cite{farzan}}, {\cite{valle}};
for earlier work see, for example, {\cite{earlySN}}. The presence of
pseudo-Goldstone bosons can affect a supernova in two different
ways. The decay of electron neutrinos into Goldstone bosons can
deleptonize the core prior to the `bounce' preventing the bounce from
taking place. This puts a bound on the electron neutrino coupling to
Goldstone bosons.  Further, after the bounce, the supernova can lose
energy too rapidly through neutrino or antineutrino decays into
Goldstone bosons, which then free-stream out. This also puts bounds on
the Goldstone boson couplings to the various neutrino species. All
these constraints tend to depend in detail on the supernova dynamics
but are typically at the $g \simlt 10^{-6}$ level or so. This is a much
weaker bound than that arising from considerations of big bang
nucleosynthesis and is therefore not included in Figure
\ref{fig:majorana}.

\subsection{\label{subsec:dirac} Dirac Neutrinos}

For Dirac neutrinos the relevant interaction vertex involves
both a left- and right-handed neutrino.
 \beq
   \mathcal{L} \supset g \nu n (H+iG)
 \eeq

The two-to-two process $\nu \bar{\nu} \leftrightarrow G G$ is changed
because now the Higgs boson couples to one left-handed neutrino and
one right-handed neutrino.  Consequently, the diagram with the virtual
Higgs boson, which was the dominant one in the Majorana case, now
needs a chirality flip for one of the initial left-handed neutrinos.
The flip suppresses this diagram by a factor of $m_\nu/T$ in the
amplitude.  The amplitude now equals that from diagrams a virtual
right-handed neutrino and no chirality flip.  Thus the rate for both
is suppressed by $m_\nu^2/T^2$ relative to Eqn.(\ref{eq:2to2rate}).
Alternatively, the process could involve an initial state right-handed
neutrino, in which case the rate will be suppressed by $r =
n_n/n_\nu$.  The value for $r$ is expected to be smaller than $0.1$
but still non-zero.  Therefore, $\nu n \leftrightarrow GG$ may
dominate over the process with the chirality flip.  As a result the
bound on $f$ is lowered to
 \beq
   f > f_4 = \left( \frac{r \sqrt{\lambda} m_\nu^2 M_{pl}}{32\pi} \right)^{1/3}.
 \eeq
The rate with the chirality flip goes like $g^4 T$ and has a recoupling 
form.  However, the region where this recoupling would lead to signals is 
already excluded by the above bound, $f>f_4$.  

The two-to-one process $\nu \nu \leftrightarrow G$ is also changed.
We must again introduce a mass insertion to convert one of the initial
neutrinos into a right-handed neutrino.  The rate will therefore be
suppressed by a factor of $m_\nu^2/T^2$ relative to the rate for the
Majorana neutrino.  Therefore, the recoupling temperature is given by
 \beq
   T_{rec} \approx \left( \frac{m_G^2 m_\nu^4 M_{pl}}{16\pi f^2}
   \right)^{1/5}
 \eeq
Demanding that $T_{rec} > m_{G}$ then results in a limit
 \beq
    f < f_1 = \left( \frac{m_\nu^4 M_{pl}}{16 \pi m_G^3} \right)^{1/2}.
 \label{eqn:nunuGrecouplimit}
 \eeq
Notice that this limit is a factor of $m_\nu/m_G$ lower than the
corresponding limit in the Majorana neutrino case
Eqn.(\ref{eqn:2to1recouplimit}). 

\begin{figure}
\begin{center}
  \includegraphics[width=16cm]{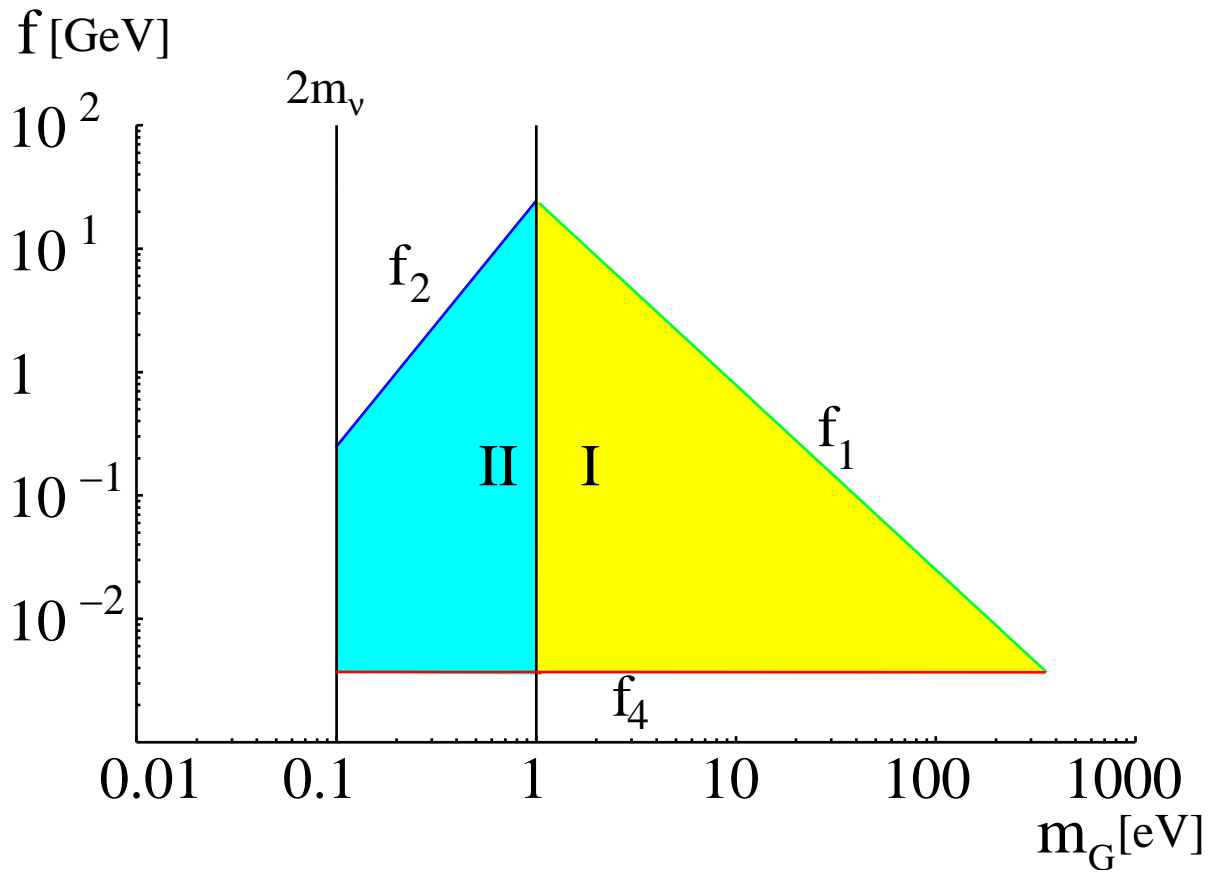}
\caption{Signal regions and cosmological bounds for a single Dirac
  neutrino coupled to a single pseudo-Goldstone boson. The region
  below line $f_4$ is exclude by BBN.  In the region above lines
  $f_1$ and $f_2$ the Goldstone boson is too weakly coupled to
  give any signal. There is a signal in $\rho_{rel}$ in region I, and
  in region II there is an overall phase shift. We have assumed $m_\nu
  = 0.05$ eV, $\lambda=1$ and $r=0.0001$. The lower bound on $f$ scales as 
  $f_4 \propto r^{1/3}$.}
 \label{fig:dirac1}
\end{center}
\end{figure}

Similarly, the bounds that must be satisfied to prevent the neutrino
 from free-streaming at $T=T_*$ and to agree with big bang
 nucleosynthesis are changed to
 \beqa
  f &<& f_2 = \left( \frac{m_G^4 m_\nu^4 M_{pl}}{16\pi T_*^7}
      \right)^{1/2} \nonumber \\
  f &>& f_3 = \left( \frac{m_G^2 m_\nu^4 M_{pl}}{16\pi T_W^5}
      \right)^{1/2}.
 \eeqa

An additional process is also possible in the case of a Dirac neutrino
mass, $\nu n \leftrightarrow G$. Again, this rate is suppressed by $r$
relative to $\nu \nu \leftrightarrow G$ in the case of Majorana
neutrinos because of the initial state right-handed neutrino. More
importantly, $n_\nu +n_n +2n_G$ is conserved.  As a result, the number
of Goldstone bosons present will never exceed $n_n =r n_\nu$.  Such a
small number of Goldstone bosons will not produce a sizable change in
the neutrino energy density.  However, if the rate is large enough, it
may still contribute to neutrino scattering thus producing a phase
shift in the angular power spectrum. For a signal to result, the
scale $f$ must be lowered by a factor of $1/\sqrt{r}$ compared to
Eqn.(\ref{eqn:2to1scatterlimit}). The rate and limit on $f$ are given by

 \beqa
   \Gamma &\approx& \frac{ m_\nu^2 m_G^2}{8\pi f^2 T} r \nonumber \\
   f &<& f_2' = \left( \frac{r m_G^4 m_\nu^2 M_{pl}}{8 \pi T_*^5}  \right)^{1/2} .
 \label{eqn:nunGrecouplimit}
 \eeqa

The signal regions and big bang nucleosynthesis bounds are shown in
figure \ref{fig:dirac1} for $m_\nu = 0.05$ eV and $r=0.0001$.  For these
values the relevant scattering process is $\nu \nu \leftrightarrow G$;
identical to the process which equilibrates the Goldstone boson.  The
signal regions are the same as those for Majorana neutrinos.  

%

\section{\label{sec:3nu} Signal Regions for Three Neutrinos}

We are now in a position to study the cosmological signals of three
neutrinos interacting with a single pseudo-Goldstone boson. As in the
single neutrino case, if recoupling occurs we expect an energy density
signal for $m_G > 1$ eV and a scattering signal for $m_G < 1$ eV.  The
crucial difference, however, is that now the number of neutrinos that
recouple to the pseudo-Goldstone boson may be one, two or three. Since
the pseudo-Goldstone boson couples to each neutrino with a strength
proportional to its mass, even at the quantum level
{\cite{neutrinostability}}, the pattern of neutrino masses determines
the number of neutrinos which recouple, and thereby the magnitudes of
the energy density and scattering signals. This is very exciting,
because it implies that a careful investigation of the cosmic
microwave background may help determine the pattern of neutrino
masses.

\begin{figure}
\begin{center}
  \includegraphics[width=16cm]{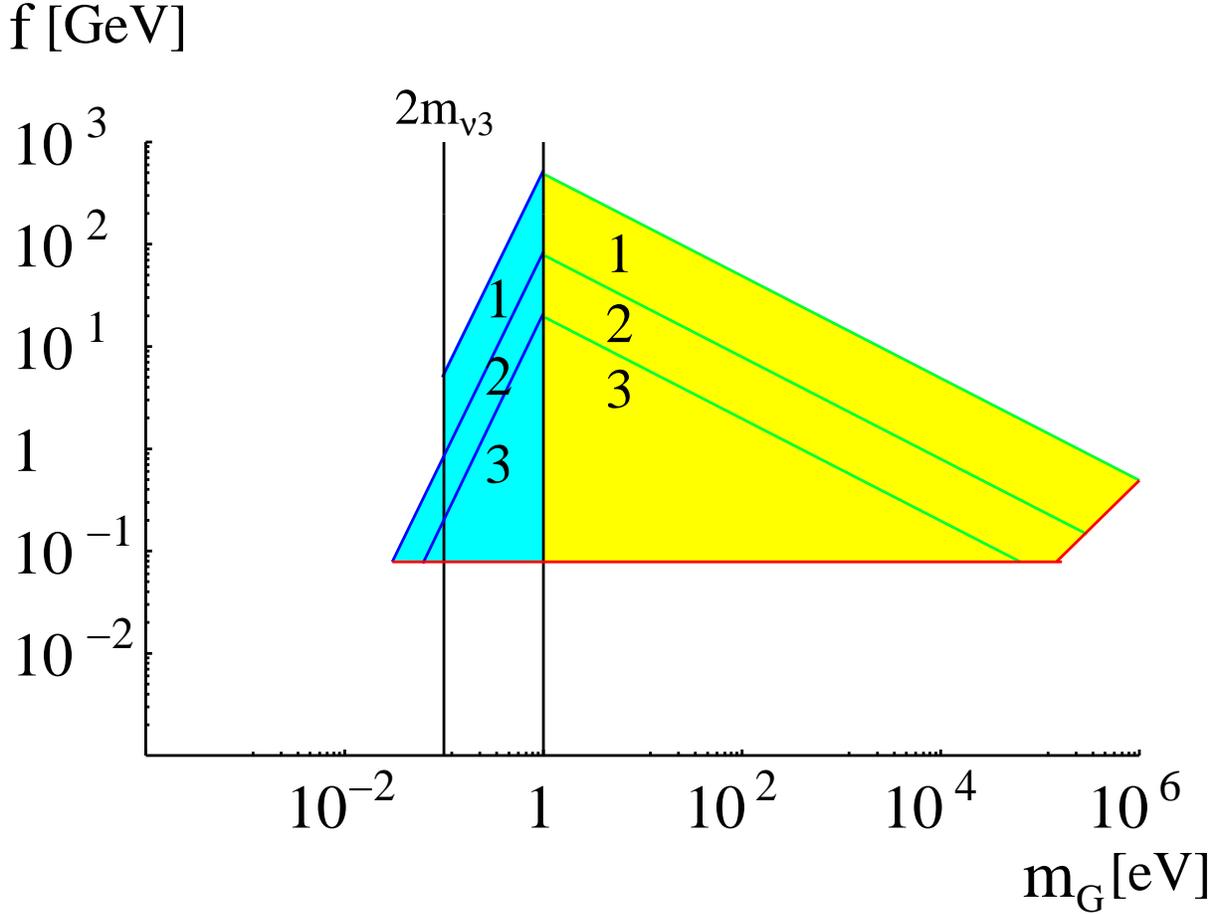}
\caption{The signal regions and bounds for three Majorana neutrinos
  with hierarchical masses $m_\nu = 0.05, 0.008, 0.002$ eV.  The
  regions are labeled by the number of neutrinos species that recouple
  to the pseudo-Goldstone boson for $m_G>1$ eV and by the number of
  neutrinos that scatter at $T \sim 1$ eV for $m_G < 1$ eV.}
 \label{fig:majorana3}
\end{center}
\end{figure}

The neutrinos may be Majorana or Dirac, the pattern of their masses
hierarchical, inverse hierarchical or degenerate. We consider each of
these cases separately, starting with the case of hierarchical
Majorana neutrinos. Oscillation data reveals that for such a pattern
the heaviest neutrino has a mass of about 0.05 eV, while the
intermediate neutrino has mass of about 0.008 eV. The mass of the
lightest neutrino is significantly smaller, and for concreteness we
take it to be 0.002 eV. The allowed signal region is plotted as a
function of the symmetry breaking scale $f$ and the the
pseudo-Goldstone boson mass $m_G$ in figure \ref{fig:majorana3}. The
various regions in this plot are generically labeled by an integer
which lies between 1 and 3. If the region lies in the $m_G > 1$ eV
portion of the plot this number indicates the number of neutrinos in
equilibrium with the heavy pseudo-Goldstone boson when it
disappeared, $n_R$. If it lies in the $m_G < 1$ eV region of the plot
it indicates the number of neutrinos scattering at an eV, $n_S$.


The hierarchy in the masses of the three neutrinos implies that there are
large, distinct parts of the plot where one, two or three neutrinos
contribute to the signal. This is to be contrasted with the case where the
spectrum of neutrino masses is inverse hierarchical or degenerate. If the
neutrino masses exhibit an inverted hierarchy, the mass of both the two
heavier neutrinos is close to 0.05 eV, with a splitting smaller than 0.001
eV. This implies that there is almost no region of parameter space in $f$
where only one neutrino contributes to a signal, and in general we expect
either two or three neutrinos to contribute. If the neutrino masses are
degenerate, then the small mass splitting (less than 0.001 eV) between each
pair of neutrinos implies that in the entire signal region all three
neutrinos will contribute. It may therefore be possible to determine the
pattern of neutrino masses from a careful measurement of the cosmic
microwave background.

How well can these different patterns be distinguished? That depends on
the magnitude of the signal for each case. The energy density signals
for a single pseudo-Goldstone boson decaying or annihilating into one, two or
three Majorana neutrinos are shown in Table \ref{table:Nnu}. We see
from the table that the differences in the energy density signal
between the various patterns is rather small, which means that it is
unlikely that we will be able to distinguish between them in upcoming
experiments.  If the PGB interacts only via $\nu \nu \leftrightarrow
G$ then the relevant signal corresponds to the column of Table
\ref{table:Nnu} labeled ``$\mu \neq 0$'', since the PGB and neutrinos possess a chemical
potential. In theories with multiple PGBs, the reaction $G'
\leftrightarrow GG$ will force the chemical potential to vanish, as we
discuss in \ref{sec:theoriesofmass},
giving larger signals, as shown in the  ``$\mu = 0$'' column.

However, the differences in the scattering signal, which
can immediately be read off from Eqn. (\ref{eq:lshiftscatt}), are
large enough that in this region of parameter space there is indeed
the distinct possibility of distinguishing between different patterns
of neutrino masses.

\begin{table}
\begin{center}
\begin{tabular}{|c||c|c||c|c|} \cline{1-5} \cline{1-5}
    &\multicolumn{2}{|c}{Dirac} & \multicolumn{2}{|c|}{Majorana} \\
 \cline{1-5} \cline{1-5}
    $n_{R}$ & $\mu = 0$ & $\mu \ne 0$ & $\mu = 0$  &
    $\mu\ne 0$  \\ \cline{1-5}
    1& 3.09 & 3.03 & 3.16 & 3.08  \\ \cline{1-5}
    2& 3.09 & 3.03 & 3.17 & 3.10  \\ \cline{1-5}
    3& 3.09 & 3.03 & 3.18 & 3.12  \\ \cline{1-5}  
\end{tabular}
\caption{Table of effective number of neutrinos as determined by the
relativistic energy density.  Predictions are given for both Dirac 
and Majorana masses and for cases in which 1, 2 or 3 neutrinos recouple
to the pseudo-Goldstone boson.}
\label{table:Nnu}
\end{center}
\end{table}

\begin{figure}
\begin{center}
  \includegraphics[width=16cm]{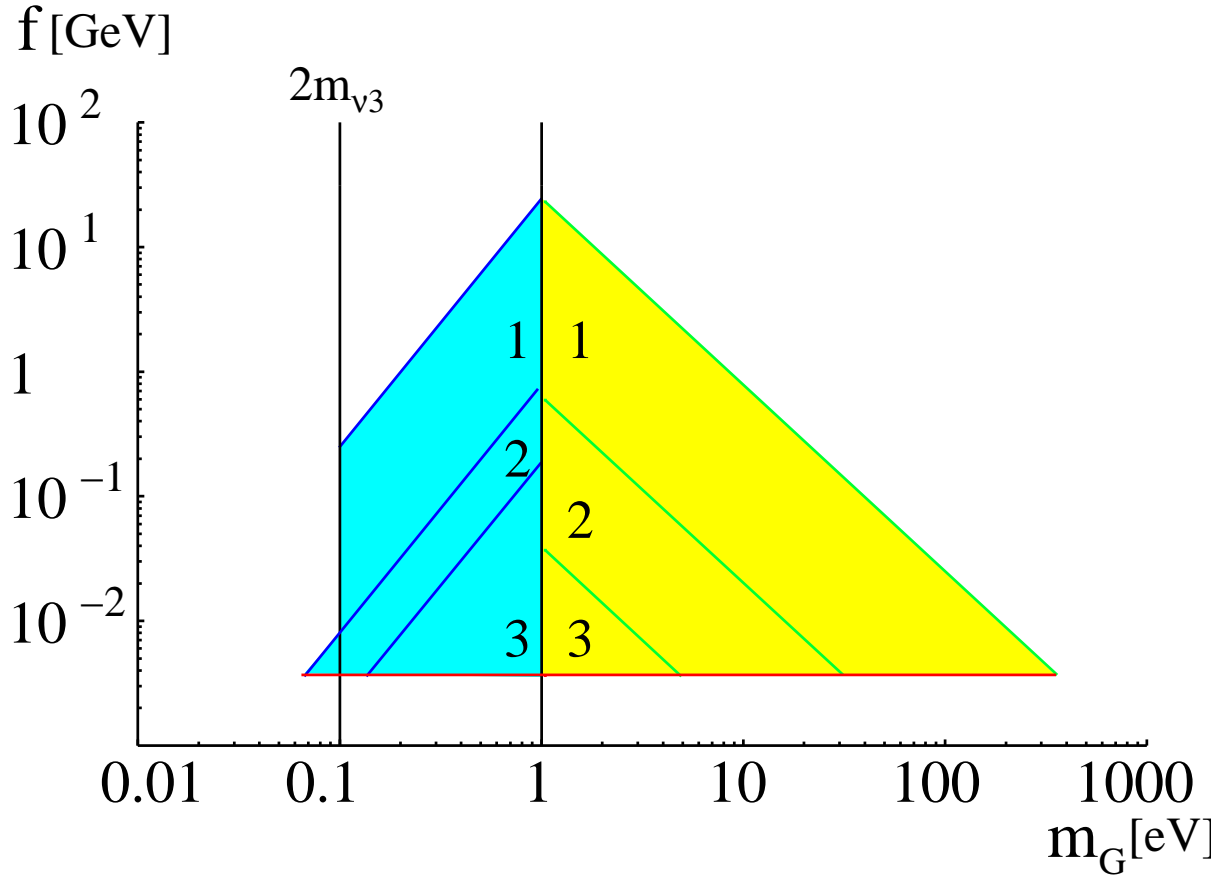}
\caption{The signal regions and bounds for three Dirac neutrinos
  with hierarchical masses $m_\nu = 0.05, 0.008, 0.002$ eV.  The
  regions are labeled by the number of neutrinos species that recouple
  to the pseudo-Goldstone boson for $m_G>1$ eV and by the number of
  neutrinos that scatter at $T \sim 1$ eV for $m_G < 1$ eV and $r=0.0001$.}
 \label{fig:dirac3}
\end{center}
\end{figure}

We now move over to the case of three Dirac neutrinos. As before we
first consider a hierarchical pattern of masses. The signal region is
as shown in figure \ref{fig:dirac3}, where we have used the same
values for the neutrino masses as in the Majorana case. As in the one
neutrino model, the signal region differs from that of Majorana
neutrinos. The reason is that, because there are a reduced number of
pseudo-Goldstone bosons and right-handed neutrinos in the bath
initially, the only $2 \leftrightarrow 1$ processes that can
significantly alter the energy density in radiation necessarily
involve a chirality flip on one of the neutrino legs, and are
therefore suppressed except at very low temperatures. The reduced
number of pseudo-Goldstones and right-handed neutrinos initially
present also weaken the bounds from BBN.  The scattering signal region
is also altered because of the reduced number of right-handed
neutrinos available for scattering.  As in the Majorana case, we
obtain either an energy density signal or a scattering signal,
depending on whether or not the pseudo-Goldstone boson mass exceeds an
eV. 

Notice that for illustrative purposes we have chosen a somewhat small
value of $r$.  As a result, for the largest mass the scattering
process is mass suppressed.  However, for the smallest mass the most
rapid scattering process is $\nu n \leftrightarrow G$ which is $r$
suppressed. (For the intermediate mass the two rates are comparable.)
As a result, the region in which there exists a scattering signal
shrinks less rapidly as $m_\nu$ is lowered from $m_2$ to $m_1$ than
the region in which $\Delta N_\nu \ne 0$. This is because the
equilibration process $\nu \nu \leftrightarrow G$ is suppressed by
$m_\nu^2/T^2$.  This leads to a discontinuity in the signal region at
$m_G=1$ eV.

Since the neutrino masses are hierarchical, there are large,
distinct regions where one, two or three neutrinos contribute to the
signal. As in the Majorana case, for an inverted hierarchy we expect
either two or three neutrinos to contribute, while for the degenerate
case all three are expected to contribute. Once again, this opens the
door to the possibility of determining the pattern of neutrino masses
from precision measurements of the cosmic microwave background.

The energy density signals for a pseudo-Goldstone boson decaying or
annihilating into one, two or three Dirac neutrinos are shown in Table
\ref{table:Nnu}, and can be contrasted with the Majorana case.  Once
again the differences in the energy density signals are too small to
distinguish between the various patterns of neutrino masses in
upcoming experiments.  However the differences in energy density
signals between the Dirac and Majorana cases are large enough that it
may be possible to distinguish between these two cases. As before, for a given
number of scattering neutrinos, the scattering signal may be
immediately obtained from Eqn. (\ref{eq:lshiftscatt}). In this region
of parameter space we expect that it will indeed be possible to
distinguish between different patterns of neutrino masses.

\section{\label{sec:leptonflavor} Multiple PGBs from
Approximate Lepton Flavor Symmetries}

Until now we have assumed that the neutrinos couple to a single 
pseudo-Goldstone boson, $G$, resulting from the spontaneous breakdown 
of an Abelian, flavor-diagonal symmetry. In the Majorana case, $G$ is the 
Majoron of broken lepton number symmetry. However, the flavor
symmetries of the neutrino sector are much richer than this, offering
the hope of larger CMB signals which could probe the mass generation
mechanism at a deeper level.

The most general lepton flavor symmetries acting on three generations
of left-handed fermions $\ell,e$ and $n$ are $U(3)_\ell \times U(3)_e
\times U(3)_n$. Here and below we give the symmetries for the Dirac
case, with the understanding that the Majorana case is trivially
obtained by deleting any symmetries on $n$. The charged lepton masses
arise from operators such as $\ell \Sigma e h$ when $\Sigma$ acquires
a vev. Part of the flavor symmetry is broken, and any Goldstones
produced at this stage are constrained to be very weakly
coupled. However, some flavor symmetries escape breaking at this stage,
and remain as symmetries of the low energy neutrino interactions. They necessarily include 
$U(3)_n$ and a flavor symmetric $U(1)$ acting on the neutrinos, which
we write as $e+ \mu + \tau = U(1)_L$. They could also include two flavor
asymmetric $U(1)s$, $e-\mu$ and $\mu - \tau$. The Goldstones which
result from the spontaneous breaking of these symmetries we call
$G_N$ because the symmetries can only be broken by Neutral lepton
masses not charged lepton masses. Some of the symmetries broken by the charged
lepton masses may also be of interest for neutrino physics. It could
be that these symmetries reappear as accidental symmetries
of the neutrino mass sector, {\it i.e.} of the interactions of
(\ref{eq:tevops}). If they are broken by vevs of $\phi$, then
Goldstones $G_C$ appear, where the label indicates that explicit
symmetry breaking arises from the Charged lepton masses.. 

We assume that all global symmetries are not exact but receive
explicit breakings from some mass scale $M_E$, which might be as high
as the Planck scale. All the Goldstones are therefore really
pseudo-Goldstones bosons, PGBs, and acquire small masses. If these arise from
dimension 5 operators, $\phi^5/M_E$, then we expect 
\begin{equation}
m_G \approx \sqrt{ \frac{f}{ M_E}} f
\approx  \mbox{eV} \left( {f \over \mbox{GeV}} \right)^{\frac{3}{2}}
 \left( {M_{Pl} \over M_E} \right)^{\frac{1}{2}},
\label{eq:mG1}
\end{equation}
where $f$ is the vev of the relevant $\phi$ field. This result is very 
interesting. All global symmetries are expected to be broken at the Planck
scale, and the simplest possibility leads to a correlation between $m_G$ 
and $f$ that passes right through our signal regions --- for both 
$\Delta N_\nu$ and $\Delta l$ signals.

This dimension 5 explicit symmetry breaking operator also induces a
low energy self interaction for the PGB, $(f/M_E) G^4$. Depending on
parameters, this may recouple the $GG \leftrightarrow GG$
reaction. However, such a process appears not to have any signals.
For the case that there are several PGBs, a much more interesting
question is whether the explicit symmetry breaking can lead to the
reaction $G' \leftrightarrow GG$ recoupling. This would have the
important consequence of forcing the chemical potential of the
Goldstones and neutrinos to vanish, and, as we have seen, this gives a
large increase to the size of the $\Delta N_\nu$ signal. We find that
the recoupling of this reaction is generic to theories of
multi-PGBs. Suppose the PGBs result from fields $\phi, \phi' ...$ and
that the explicit symmetry breaking at dimension 5 includes several
operators $(1/M_E)(a \phi^5 + b \phi^4 \phi' + c \phi^3 \phi'^2 +
...)$. Substituting $\phi = (f + iG)/\sqrt{2}, \phi' = (f' + i
G')/\sqrt{2}, ...$, and performing field phase redefinitions to remove
linear terms in $G,G' ...$, one discovers that the phases of $a,b,c,
... $ are not all removed, so that interactions of the form $G'GG$
appear in the low energy theory, even after rotating to the PGB mass
basis.  The recoupling temperature for $G' \leftrightarrow GG$ is
\begin{equation}
T_{rec}(G' \leftrightarrow GG) \approx 
\left( {m_{G'}^4 M_{pl} \over f^2 16\pi} \right)^{1/3}
\approx \mbox{keV} \left( {m_{G'} \over \mbox{keV}} \right)^{4/3}
\left( {\mbox{GeV} \over f} \right)^{2/3},
\label{eq:G'GG}
\end{equation}
where $f$ is the larger of $f$ and $f'$, and we have taken $G'$
heavier than $G$. For the entire region of $f$ and $m_G$ of interest for the
$\Delta N_\nu$ signal, $T_{rec} > m_{G'}$. Hence, for multi-PGB
theories, the generic situation is that $G'
\leftrightarrow GG$ recouples and we expect the larger $\Delta N_\nu$ 
signals appropriate for vanishing chemical potential.

The difference
between the $G_N$ and $G_C$ PGBs, is that the $G_C$ also have
contributions to their mass from explicit breakings of the accidental
low energy neutrino symmetries by the interactions that generate the
charged fermion masses. Although, these contributions are model dependent, 
for a wide range of theories we can estimate their size. Suppose that the 
explicit symmetry breaking appears in the low energy neutrino theory as 
powers of the insertion $\lambda_E$, the charged lepton Yukawa coupling
matrix, which transforms under $SU(3)_\ell \times SU(3)_e$ as $(3,3)$.
Assume also that, in the symmetric limit, a potential for $\phi$ is generated 
with a mass term of order $-f^2 \phi^\dagger \phi$. Including the effects of 
explicit symmetry breaking, this term will be modified to
$-f^2 \phi^\dagger (1 + c \lambda_E \lambda_E^\dagger) \phi$, 
where $c$ depends on coupling parameters and loop factors. Ignoring $c$, 
gives
\begin{equation}
m_{G_C} \approx \lambda'_E f
\approx ( \mbox{keV -- $10$ MeV} ) {f \over \mbox{GeV}},
\label{eq:mG2}
\end{equation}
where $\lambda'_E$ are the eigenvalues of the charged lepton Yukawa
coupling matrix. The $G_C$ PGBs
have hierarchical masses typically in the range to give a
$\Delta N_\nu$ CMB signals, while $G_N$ are typically lighter and
could give either a  $\Delta N_\nu$ or a $\Delta l$ CMB signal.

In Table \ref{table:G} we list the number of each type of PGB, for both
Majorana and Dirac cases, when the flavor symmetry is maximal. 
It is immediately clear that if there are
anywhere near these numbers of PGB, then the $\Delta N_\nu$ signal is
likely to be much larger than in the previous sections. How many
Goldstones do we expect? Zero, 1 or of order 10? 
If the mass scale of the physics generating neutrino masses is
substantially below the weak scale then it is plausible that this
scale, $f$, results from the breaking of a global symmetry. There
need only be one symmetry, leading to a single PGB as discussed in
the previous sections. However, if the entire structure of the
neutrino mass matrix follows from broken symmetries, in analogy to the
Froggatt-Nielson symmetry breaking frequently studied in the charged
sector, then many more Goldstones are to be expected. While the number
need not necessarily be the maximal number listed in Table \ref{table:G}, no
non-Abelian subgroup can escape breaking, since we know that the three
neutrinos have no exact degeneracies, and hence, if the underlying
symmetries of the neutrino sector is $U(3)_\ell \times U(3)_n$, 
we expect the number of Goldstones to be near maximal.

\begin{table}
\begin{tabular}{|c||c|c|}
\hline
      &  Majorana  & Dirac \\
\hline \hline
$G_N$ &  1 -- 3    &  10 -- 12 \\
\hline
$G_C$ & 8 -- 6 & 8 -- 6\\
\hline
\end{tabular}
\caption{The number of $G_N$ and $G_C$ PGBs in theories with
  three generations of Majorana or Dirac neutrinos and maximal flavor 
symmetry. For the Majorana (Dirac) case there are a total of 9 (18) PGB.
The number of $G_C$ is increased from 6 to 7 or 8 if the physics 
responsible for the charged lepton masses breaks $e-\mu$ and 
$\mu- \tau$ symmetries.}
\label{table:G}
\end{table}

As an example, consider a $U(3)_\ell$ theory of Majorana neutrinos
based on operators of the form $ g \nu_i \phi_{ij} \nu_j$ where the
generation indices $i,j$ run over 1,2,3.  Suppose that the vevs of the $\phi$ 
multiplet are hierarchical, and we have chosen the basis where the largest vev
lies in the 33 direction, breaking
$U(3)_\ell$ to $U(2)_\ell$, creating 5 PGB. Since 4 of these PGBs are
associated with off-diagonal generators, they are $G_C$. These 5 PGB
all have the same coupling, $g$, to neutrinos and for a large range of
parameters will be produced by the $2 \leftrightarrow 1$ process
$\nu_i \nu_j \leftrightarrow G_{ij}$. Other $\phi$ components with smaller 
vevs will break the remaining $U(2)_\ell$ symmetry
creating further PGBs. Since the majority of PGBs are $G_C$ rather 
than $G_N$, they are likely to have hierarchical masses, for example
as in (\ref{eq:mG2}), that are heavy enough to give a CMB $\Delta N_\nu$
signal. A similar analysis
would occur for a $U(3)_\ell \times U(3)_n$ theory of Dirac
neutrinos, with two important differences. First, many more $G_N$ are
expected from the $U(3)_n$ symmetry. 
However, countering this, for any Goldstone with dominant
coupling to mass eigenstates $\nu_i n_j$ the rate for being recoupled
by the $2 \leftrightarrow 1$ process is suppressed by a factor
$(m_j/T)^2$ at temperature $T$. This would make the $\Delta N_\nu$ signal
sensitive to the spectrum of both  the neutrinos and the PGBs.

Rather than compute the spectrum for any specific model, in Table 
\ref{table:manyG} we present the energy density signal for generic situations. 
A particular model is likely
to have a signal which can be computed and differs from those in the
Table. However, the Table does give an impression of the size of the
signals which should be expected.
For illustration we have taken Majorana models with 2 or 8 PGB that 
have recoupled to 1,2 or 3 neutrinos and are heavier than 1 eV. This 
could be the situation for models based on $U(1)_e \times U(1)_\mu
\times U(1)_\tau$ or $U(3)_\ell$, where only the PGB for overall lepton
number is lighter than 1 eV. For Dirac neutrinos we have chosen  2, 8
or 16 heavy PGB, corresponding to models based on $U(1)_e \times U(1)_\mu
\times U(1)_\tau$, $U(3)_{\ell+n}$ or  $U(3)_\ell \times U(3)_n$
symmetries, where again one or two flavor diagonal PGB  are taken
lighter than 1 eV. We note that in both Majorana and Dirac cases,
planned CMB experiments are now able to distinguish between
cases where the PGB recouples to 1,2 and 3 neutrinos, giving
sensitivity to differing neutrino spectra. The energy density signal
clearly grows substantially when more PGB contribute. 

\begin{table}
\begin{center}
\begin{tabular}{|c||c|c||c|c|} \cline{1-5} \cline{1-5}
    &\multicolumn{2}{|c}{Dirac} & \multicolumn{2}{|c|}{Majorana} \\
 \cline{1-5}
    $n_{G}$ & $\mu = 0$ & $\mu \ne 0$ & $\mu = 0$  &
    $\mu\ne 0$  \\ \cline{1-5}
    2& 3.18 & 3.06 & 3.34 & 3.19  \\ \cline{1-5}
    8& 3.62 & 3.18 & 4.08 & 3.42  \\ \cline{1-5}
    16& 4.08 & 3.27 & X & X  \\ \cline{1-5}
\end{tabular}
\caption{Table of effective number of neutrinos as determined by the
relativistic energy density.  Predictions are given for both Dirac and
Majorana masses and for cases in which $n_G$ Goldstone bosons recouple
to $n_R=3$ neutrinos and then decay.  We have assumed that all
Goldstone bosons equilibrate prior to any decaying. We assume that all
heavy Goldstone bosons have decayed or annihilated prior to the
equilibration of any light PGBs.}
\label{table:manyG}
\end{center}
\end{table}

\begin{table}
\begin{center}
\begin{tabular}{|c||c|c||c|c|} \cline{1-5} \cline{1-5}
    &\multicolumn{2}{|c}{Dirac} & \multicolumn{2}{|c|}{Majorana} \\
 \cline{1-5}
    $n_R$ & $\mu = 0$ & $\mu \ne 0$ & $\mu = 0$  &
    $\mu\ne 0$  \\ \cline{1-5}
    3& 3.62 & 3.18 & 4.08 & 3.42  \\ \cline{1-5}
    2& 3.58 & 3.15 & 3.97 & 3.35  \\ \cline{1-5}
    1& 3.49 & 3.12 & 3.77 & 3.24  \\ \cline{1-5}
\end{tabular}
\caption{Table of effective number of neutrinos as determined by the
relativistic energy density.  Predictions are given for both Dirac and
Majorana masses and for cases in which $n_G=8$ Goldstone bosons
recouple to $n_R$ neutrinos and then decay.  We have assumed that
all Goldstone bosons equilibrate prior to any decaying.  We assume
that all heavy Goldstone bosons have decayed or annihilated prior to
the equilibration of any light PGBs.}
\label{table:nrec}
\end{center}
\end{table}

In the Dirac case one might question whether it is really plausible
that so many PGB are recoupled --- as mentioned earlier, a PGB $G_A$ that
couples predominantly to neutrino mass eigenstates via
$\nu_i T^A_{ij} n_j$ has a recoupling rate suppressed
by $(m_j/T)^2$. However, the issue here is whether the mass eigenstate
bases for the PGB and the neutrinos line up. Since the explicit
symmetry breaking from very high scales, or from the charged lepton
mass sector, is unrelated to the neutrino mass generation, it seems
that the two bases will be very different. Indeed, for the symmetry 
breaking coming from the charged lepton sector, neutrino oscillation
data already tell us that there are large mixing angles between the
two bases. Hence, if we consider the coupling to the
heaviest mass eigenstate neutrino $\nu_3 n_3 G_{33}$, we expect that
$G_{33}$ will be a linear combination of all the PGB mass eigenstates
without any small mixing angles. Hence the suppression factor for all
of them is only $(m_3/T)^2$. If one of the PGB recouples we typically
expect that they all do. 

This does not mean that the signal is insensitive to the neutrino
spectrum, since the signal does depend on the number of neutrinos that
get recoupled. From Table \ref{table:nrec} we see that the dependence
on the number of recoupled neutrinos is large and future CMB
experiments will be able to probe the nature of the neutrino spectrum,
unlike the case of a single PGB discussed in the previous section.

The change in the neutrino energy density is now so large that this
has repercussions for the size of the scattering signal.  This results
from large deviations of $N_\nu$ from three in
Eqn. (\ref{eq:lshiftdeltaE}).  Additionally, if some neutrinos that
were coupled to the heavy pseudo-Goldstone bosons are not coupled to
the light PGB and hence free-stream, it is possible that $N_{\nu FS}
\ne (3-n_S)$.  In general, assuming that all scattering neutrinos and
light Goldstone bosons were heated by the disappearance of the heavy
PGBs, $N_{\nu FS}$ is given by
 \beq
   N_{\nu FS} = \left( 3-n_R \right)+ \frac{n_R - n_S}{n_R + g_G / g_\nu} \left( N_\nu -
   \left( 3-n_R \right) \right) 
 \eeq  
where $g_\nu = 7/4$ or $7/2$ if the neutrinos are Majorana or Dirac
and $g_G$ is the spin degeneracy of the light PGBs.  This, combined
with Eqn. (\ref{eq:lshiftdeltaE}) allows us to calculate $\Delta l_n$.

\begin{table}
\begin{center}
\begin{tabular}{|c|c|c|} \cline{1-3} 
    $n_R$ & $N_{\nu FS}$ & $\Delta l_n$ \\
 \cline{1-3} \cline{1-3}
    1& 2.00 & -14.04 \\ \cline{1-3}
    2& 2.16 & -14.77 \\ \cline{1-3}
    3& 2.28 & -15.45 \\ \cline{1-3}  
\end{tabular}
\caption{Illustration of how the scattering signal,  $\Delta l_n$, depends on
  $\Delta N_\nu$ and  $\Delta N_{\nu FS}$. These predictions are for the case that
  one Majorana neutrino scatters due to interactions with one 
  light pseudo-Goldstone boson. In this illustration,  $\Delta N_\nu$ and 
  $\Delta N_{\nu FS}$ are calculated by assuming that
  $n_R$ neutrinos, including the one that scatters, are 
  heated by the disappearance of 8 heavy PGBs. These  $\Delta l_n$ predictions
  should be compared to the value $-15.6$ that results for $N_\nu =3$ and  
$N_{\nu FS}=2$.}
\label{table:deltal}
\end{center}
\end{table}

Notice that $\Delta l_n$ is no longer proportional to $n_S$ and in fact
has a dependence on $n_R$ and $n_G$ through $N_{\nu FS}$ and
$N_\nu$.  As a result, the phase of the acoustic oscillations may
give us information about the number of neutrino species that were
coupled to the heavy pseudo-Goldstone bosons and the number of heavy
PGBs as well as the number of neutrinos that are scattering at an eV.
An example of this is shown in Table \ref{table:deltal}, which shows
the case of Majorana neutrinos coupled to 8
heavy and 1 light PGB.  We assume that one neutrino
species is scattering and that this species was also heated by the
decay or annihilation of the heavy PGBs.  We see that in this case the
$\Delta l_n$ shift has a noticeable dependence on the number of recoupled
neutrino species, in contrast with Eqn. (\ref{eq:lshiftscatt}), and 
could be discovered with an observational resolution of  
$\Delta l_n \leq \pm 1$.

\section{\label{sec:theoriesofmass} Probing Theories of Neutrino Mass}

One aspect of the PGB couplings is very general, and is
independent of the underlying symmetry structure, the $\phi$ multiplet
structure and the interactions coupling $\phi$ to neutrinos.
If a PGB is produced at a symmetry breaking of strength $f$ that
produces a mass term $m_\nu$ for some neutrinos, then the
PGB coupling strength to these neutrinos is $g=m_\nu/f$. This means
that the analysis of section \ref{sec:1nu} applies to any individual PGB,
providing $m_\nu$ is interpreted as the neutrino mass arising from
this particular PGB coupling. For the Majorana case, the lines of
Figure 1 do not depend on $m_\nu$, hence the only modification necessary
is to rescale the $f$ axis by a factor $m_\nu/0.05$eV. From this
we see that CMB signals may arise in {\em any} theory of Majorana
neutrinos where neutrino flavor symmetries are broken in the mass range
\begin{equation}
f = ( \mbox{ $50$ MeV -- $500$ GeV}) \frac{m_\nu }{ 0.05 eV}.
\label{eq:frange}
\end{equation}
For the Dirac case, the slopes of some of the lines do depend on
$m_\nu$, and some limits depend on $r$, so that the relevant regions
are shifted to somewhat lower values of $f$. 
We seek theories where lepton flavor symmetries are broken at the 
weak scale, or up to four orders of magnitude below the weak scale. 
Whatever breaks the weak interaction might also break lepton flavor 
symmetries, either directly or at one or two loop order.

This large range in $f$ 
allows a wide variety of models, not just those of
Eqn. (\ref{eq:tevops}), but does not include the popular seesaw models
where lepton number is broken quite close to the scale of grand
unification.  In general two factors contribute to explaining why the
neutrinos are much lighter than the weak scale: $f/v$ and $v/M$. The
relative importance of these two factors is model dependent. However,
for all models it is clear that a crucial ingredient is that an
approximate lepton flavor symmetry is broken at some scale $f$ much
less than the energy scale, $M$, responsible for the physics of
neutrino masses.  Below we choose two models to illustrate the rich
range of possibilities.

\subsection{A low energy seesaw}

It may be that $M \approx v$ and the suppression of neutrino masses is
entirely due to a small value for $f/v$. As an example of such a
theory, consider the seesaw model:
\begin{equation}
L = nn \phi + \ell nh  \left( \frac{\phi'}{ v} \right)^2,
\label{eq:seesawL}
\end{equation}
where order unity couplings are understood. Suppose that lepton number
is spontaneously broken at the weak scale $\phi \approx v e^{iG/v}$,
while other lepton flavor symmetries are broken at some lower scale
$f$:  $\phi' \approx f e^{iG'/f}$. The light neutrinos are Majorana with
a mass seesawed down to
$f^4/v^3$, so that $f$ should be of order 1 GeV. There are two very
different types of PGB: $G$ and $G'$ with symmetry breaking parameters
$v$ and $f$. From Figure \ref{fig:majorana} we find that either could give scattering
or energy density CMB signals. The mass of $G$ would need to be
close to 1 eV, but $G'$ could give a signal for a wide range of
masses. 

\subsection{A $SU(3) \times U(1)_L$ theory}

A discussion of realistic three neutrino theories with a
single Goldstone was given in section \ref{sec:3nu}. With more than one
Goldstone an important new ingredient appears: multiple symmetry
breaking scales $f$. 
Neutrino mass ratios may now arise from a hierarchy of $g$ parameters 
or from a hierarchy of $f$ parameters.  Of course, given the observed 
pattern of neutrino oscillations the hierarchies need not be large.
In the case of hierarchical flavor 
symmetry breaking, for a fixed pattern of neutrino masses, smaller 
values of $f$ translate into larger values for $g$. Hence the couplings 
of the PGBs to the lighter neutrinos are larger than for the case of a single 
symmetry breaking scale, discussed in section \ref{sec:3nu}, and the signal regions are 
correspondingly enhanced for the lighter neutrinos compared to the 
regions shown in Figures \ref{fig:majorana3} and \ref{fig:dirac3}.

A particularly interesting model with a hierarchy of global symmetry 
breaking scales is a $SU(3) \times U(1)_L$ theory of Majorana neutrinos
described by 
\begin{equation}
L =  \frac{ 1}{ M^3} \ell_i \left( \phi_{A_{ij}} \phi_L \right) \ell_j hh
\label{eq:31model}
\end{equation}
where $\phi_L$ carries overall lepton number, while the multiplet
$\phi_A$ is a representation of $SU(3)$. This theory not only has
multiple symmetry breaking scales with hierarchical vevs of $\phi_A$, but 
neutrino masses occur at second order in symmetry breaking, rather than at 
linear order.

The neutrino mass is a product of symmetry breaking terms
\begin{equation}
m_\nu =  \frac{f_A f_L}{ M}  \frac{v^2}{ M^2},
\label{eq:31mnu}
\end{equation}
that could lead, for example, to a hierarchical spectrum. The coupling
of the Goldstone of overall lepton number, $G_L$,  is proportional to
the $SU(3)$ symmetry breaking:
\begin{equation}
g_L =  \frac{f_A }{ M}  \frac{v^2}{ M^2},
\label{eq:gL}
\end{equation}
while the couplings of the Goldstones of broken $SU(3)$ are
proportional to the breaking of lepton number:
\begin{equation}
g_A =  \frac{f_L}{ M} \frac{v^2 }{ M^2}.
\label{eq:gA}
\end{equation}

The CMB signals for this model vary drastically as $(f_A,f_L)$ are
varied.  If $f_{A,L} > v$ there are no CMB signals. If only $f_L < v$
then there is a single flavor-diagonal PGB, $G_L$, with signals as
given in section \ref{sec:3nu}.  On the other hand if $f_A < v$ while
$f_L > v$, then there are several PGB, $G_A$, which contribute to CMB
signals and which may have a hierarchy of $f_A$. If $f_{A,L} <v$ we
have a model of Majorana neutrinos where the maximal number of PGBs
can contribute to CMB signals. Since $G_L$ is a member of $G_N$, it
may have a small mass and give rise to a scattering signal. Even
though it derives from overall lepton number symmetry, it has a
different coupling to each neutrino species, so that the scattering
signal may correspond to 1,2 or 3 neutrinos depending on the neutrino
spectrum.  The 8 flavor PGB, $G_A$, may all be $G_C$ type with larger
masses, giving a very large $\Delta N_{\nu}$ signal as shown in Table
\ref{table:manyG}.

\section{$f<v$ from Supersymmetry}
\label{sec:susy}

CMB signals are possible even if the lepton flavor symmetries are
broken at scales as high as the electroweak scale, $v$. However,
observable signals result from a much wider range of PGB masses if $f
< v$. There are a variety of scenarios for naturally inducing symmetry
breaking scales well beneath the weak scale; in this section we study
a supersymmetric theory.

Consider the superpotential below, where in addition to a Dirac mass for the neutrino of the form
of equation (\ref{eq:tevops}), we have added a mass $M_N$ for the right-handed neutrino. As we
will argue later, the natural size of $M_N$ is of order the weak scale, and therefore the neutrino
masses in this theory are Majorana. The theory has an R symmetry corresponding to lepton number
under which $L, N$ and $\Phi$ are all charged. In the absence of the mass term $M_N$ for the
right-handed neutrino the theory has an additional right-handed lepton number symmetry under
only $N$ and $\Phi$ are charged; this symmetry is broken at the weak scale generating a
right-handed neutrino mass.

\begin{eqnarray}
  \mbox{W} = \frac{\lambda}{M} L N H_u \Phi + M_N N^2
               + \frac{\alpha}{3!} \Phi^3
\label{eq:superpot} ~.
\end{eqnarray}
After supersymmetry and electroweak symmetry breaking, we obtain the
coupling $\sqrt{2} g_{ij} \nu_i n_i \phi$ which leads to a Dirac mass term,
and the scalar potential
\begin{eqnarray}
   V =
                  \tilde{m}^2 (
                  |\tilde{\nu}_i|^2 + |\tilde{n}_i|^2 )
                + | g_{ij} \tilde{\nu}_j \phi + 
                 2 \left({M_N} \right)_{ij} \tilde{n}_j|^2
                + | g_{ij}\tilde{n}_j|^2 |\phi|^2
                + |\frac{\alpha}{2}\phi^2 + g_{ij}\tilde{\nu}_i \tilde{n}_j|^2     ~,
\label{eq:lag}
\end{eqnarray}
where we have taken a common soft mass, $\tilde{m}$,
for $\tilde{\nu}_i$ and $\tilde{n}_i$ for simplicity.  Note that
we have made a crucial assumption that $\Phi$ does not
feel supersymmetry breaking directly; namely there is no
soft mass for $\phi$.  This would occur, for example, in certain
theories of gauge-mediated supersymmetry breaking if
$\Phi$ does not have any  gauge interactions. Since $\Phi$ couples
through $g_i$ to particles which do feel supersymmetry breaking,
radiative corrections must induce a supersymmetry breaking mass at
least as big as
\begin{eqnarray}
   \delta m_\phi^2 = -\frac{g^2}{16 \pi^2} \tilde{m}^2,
\end{eqnarray}
where $g$ is the largest coupling of $\Phi$ to neutrinos.
Note that the sign is negative, and will induce symmetry breaking, in
a very similar fashion to radiative electroweak symmetry breaking in
the standard supersymmetric model.
Therefore, the VEV of $\phi$, namely $f$, is given by
\begin{eqnarray}
  f \simeq \sqrt{\frac{-2\delta m_\phi^2}{\alpha^2}}
    \approx \frac{g\tilde{m}}{4 \pi\alpha}.
\label{f1}
\end{eqnarray}
For $\alpha$ of order unity, we see that supersymmetry protects $f$
down to the scale $f_{susy} \approx g \tilde{m} / 4 \pi$. If there are
several $\Phi$ multiplets with a hierarchy of couplings to neutrinos,
then their symmetry breaking scales will reflect that hierarchy. The
coupling $g$ is naturally small, $g \approx v/M$, and is related to
the observed neutrino mass as
\begin{equation}
m_\nu \approx g^2 \frac{f^2}{M_N}
\label{f2}
\end{equation}
Eliminating $g$ between Eqs. ({\ref{f1}}) and ({\ref{f2}}) we
obtain an approximate expression for the scale $f$ given below
\begin{equation}
f^2_{susy}   \approx   \frac{\tilde{m}}{ 4 \pi} \sqrt{m_{\nu} M_N}
\label{eq:fsusy}
\end{equation}
A natural value for the scale $M_N$ at which right-handed lepton
number is broken is the weak scale. This scale could for example
be generated from a
term of the form $\lambda_N \Phi_n N N$ in the superpotential. Here $\phi_n$ 
acquires a VEV and thereby breaks right-handed lepton
number, and $\lambda_N$ is a coupling constant. If $\lambda_N$ is of
order one and the $\tilde{n}$'s have a soft mass of order the weak scale
then $S$ can acquire a VEV of order the weak scale radiatively, exactly
the way the Higgs does in the minimal supersymmetric model.

For $m_\nu = 10^{-2}$ eV and $\tilde{m} = 100$ GeV we find that
supersymmetry is able to protect $f$ down to 30 MeV. This is the
lowest $f$ for which our CMB signals are compatible with BBN
constraints. Larger $f$ could be obtained in this simple model of
radiative lepton flavor symmetry breaking 
in several ways, for example by reducing the coupling $\alpha$
or by introducing additional couplings of $\Phi$ to the supersymmetry
breaking sector. Hence we conclude that no fine tuning is necessary
anywhere in the wide range of $(f,m_G)$ parameter space that leads to
CMB signals.

\section{Conclusions}
\label{sec:conc}

We have proposed that future measurements of the CMB  will provide a
powerful probe of theories of neutrino mass that have lepton flavor
symmetries spontaneously broken at or below the weak scale. Such
theories lead to light pseudo-Goldstone bosons that interact with
neutrinos with couplings proportional to the neutrino masses. Such
interactions can modify the acoustic oscillations of the electron-photon
fluid during the eV era. In particular there is a change in the
relativistic energy density, parameterized by an effective change in
the number of neutrino species, $\Delta N_\nu$, and a change in the
multipole of the $n$th CMB peak, $\Delta l_n$, for large $n$. While
other new physics could lead to an energy density signal, a
uniform shift in the high $n$ peaks to larger $l$ can only result from
scattering of the known neutrinos, and is therefore an ideal test of
our class of theories.

The present experimental limit on deviations from the relativistic
energy density predicted by the standard model is roughly
$-2 < \Delta N_\nu < 4$ \cite{BBN} \cite{Hannestad:2003xv}, 
although the precise numbers depend 
what other data are included in the fit. Our predictions are well within
these bounds, typically in the range $0 < \Delta N_\nu < 1$. The
Planck experiment is expected to reach a sensitivity of $\pm 0.20$ at one
standard deviation, and the proposed satellite experiment CMBPOL could
probe to $\pm 0.05$ \cite{BS}. These projections assume that the
neutrinos are free-streaming; in the presence of our scattering signal
we do not know how well the energy density can be measured.
The inherent limit on  $\Delta N_\nu$ from cosmic variance is at the 
level of $\pm 0.04$ for measurements up to $l$ of 2000, if the CMB is
to determine all the cosmological parameters \cite{cosvarlim}. If the CMB is used only
to determine $N_\nu$, then the cosmic variance limit drops to $\pm
0.003$. All numbers assume polarization correlations are measured as
well as the temperature correlations. While the positions of the low
$n$ CMB peaks are now accurately determined, the higher $n$ peaks will
have to await future measurements at higher $l$. To first
approximation, our signal is $\Delta l_n \approx +8 n_S$, where $n_S =
1,2,3$ is the number of neutrino species that scatter by PGB exchange
during the eV era. This is a clear order of magnitude larger than the 
expected reach of Planck,  $\Delta l \approx \pm 2$, and CMBPOL,  $\Delta
l \approx \pm 1$, always at 1 standard deviation \cite{BS}. 

The signals can be precisely calculated in any particular theory of
neutrino masses, and reflect both the spectrum of the neutrinos and
whether the neutrinos are Majorana or Dirac. Signals are expected for
a wide range of lepton symmetry breaking parameters, 10 MeV  $<f<$ 
TeV, and for a wide range of the PGB masses, $m_G < $ MeV. The signal
regions for a Goldstone boson produced at scale $f$ with mass $m_G$
are shown in Figure 1 for Majorana neutrinos and in Figure 2 for Dirac
neutrinos. In the case that the same symmetry breaking scale $f$
produces all three neutrino masses, the signal regions are shown in
Figures 3 and 4, for a hierarchical pattern of neutrino masses. The
size of the signals differs in the regions labeled 1,2,3, and the
sizes of these regions change as the pattern of neutrino masses
changes. 

In many cases the scattering signal depends only on the number of
scattering neutrino species and is given by  $\Delta l \approx +7.8
n_S (\Delta l_{peak}/300)$, relative to the prediction of the standard model, 
which will be easily seen in upcoming
experiments. On the other hand, the size of  $\Delta N_\nu$ is highly
dependent on the number and spectrum of PGBs and the neutrino
spectrum. For a single PGB the signal is small; for example, with a
non-zero chemical potential  $\Delta N_\nu = 0.03, 0.10$ for the
Dirac, Majorana cases, for the PGB
recoupling to $n_R = 2$ neutrino species. The dependence on $n_R$ is
mild, so that in this case the signal does not allow a discrimination
between hierarchical, inverted or degenerate spectra.

The size of the  $\Delta N_\nu$ signal increases dramatically in
theories with multiple PGB. This is partly due to the larger number of
degrees of freedom, and partly because the interactions between
different PGB generically sets the chemical potential to zero. 
For example, for 8 PGB, corresponding to breaking a $SU(3)$ lepton
flavor symmetry, if all three neutrino are recoupled, $n_R = 3$,
then  $\Delta N_\nu = 0.62,1.08$ for the Dirac and Majorana cases.
The signals for $n_R = 1,2,3$ are $0.49, 0.58,0.62$ for the Dirac case
and $0.77,0.97,1.08$ for the Majorana case. Hence a precise
measurement of  $\Delta N_\nu$ has the ability to distinguish 
both the type and spectrum of the neutrinos. In cases such as this,
where there is a large  $\Delta N_\nu$ signal, there are deviations
from the simple prediction for the scattering signal, $\Delta l_n$,
which now also depends on $n_R$ and the number of PGB.

For a theory of neutrino masses to give a CMB signal, the crucial
ingredient is the spontaneous breaking of lepton flavor symmetry at a
scale $f$ of the weak scale or below. This does not occur in the 
conventional seesaw models. We have shown that the well-known
radiative symmetry breaking mechanism of supersymmetry can yield such
values for $f$, and we have demonstrated that the explicit
breaking of the global lepton symmetries expected  from the Planck
scale leads to PGB masses precisely in the range that gives CMB signals.  
There is a rich variety of models with CMB signals. We have given two
illustrations where the interactions of the neutrinos are bilinear in
the field $\phi$ that breaks the lepton symmetries. In the first model
the right-handed neutrinos are at the weak scale, $v$, and the lightness of
the neutrinos is due to powers of $f/v$. In the second model there are
two types of PGB, the flavor-diagonal Majoron,  and the flavor-changing 
PGB associated with $SU(3)$. One possibility is that the
latter contribute to a large  $\Delta N_\nu$ signal, while the Majoron 
is lighter and gives a scattering signal. It is remarkable that the
CMB offers a powerful probe of this physics.

\section*{Acknowledgments}
\label{acknowledgements}

We thank Maxim Perelstein and Martin White for valuable discussions and 
Manoj Kaplinghat, Lloyd Knox, Uros Seljak and David Spergel for communications.
T.O. thanks Danny Marfatia for many useful conversations.
This work was supported in part by the Director, 
Office of Science, Office of High Energy and Nuclear Physics, of the
U.S. Department of Energy under Contract DE-AC03-76SF00098 
and DE-FG03-91ER-40676, and in
part by the National Science Foundation under grant PHY-00-98840.

\end{document}